\documentclass{aa}  

\usepackage{graphicx}
\usepackage{txfonts}
\usepackage{etoolbox}
\usepackage{multirow}
\usepackage{longtable}
\usepackage{pdflscape}
\usepackage{wasysym}
\usepackage{amssymb}
\usepackage{array}

\newcommand{\mytilde}{\raise.17ex\hbox{$\scriptstyle\mathtt{\sim}$}}

\begin{document}

   \title{One of the closest exoplanet pairs to the 3:2 mean motion resonance: K2-19b \& c\thanks{Using observations made with SOPHIE on the 1.93-m telescope at Observatoire de Haute-Provence (CNRS), France.}
}

   \subtitle{}
   \titlerunning{K2-19b \& c}
   \authorrunning{Armstrong et al.}
   \author{David J. Armstrong\inst{\ref{inst1}} \and Alexandre Santerne\inst{\ref{inst5},\ref{inst2}} \and Dimitri Veras\inst{\ref{inst1}}
          \and Susana C. C. Barros\inst{\ref{inst5}} \and Olivier Demangeon\inst{\ref{inst5}} \and Jorge Lillo-Box\inst{\ref{inst11}} \and James McCormac\inst{\ref{inst1}}
           \and Hugh P. Osborn\inst{\ref{inst1}} 
            \and Maria Tsantaki\inst{\ref{inst2}}  
           \and Jos\'e-Manuel Almenara\inst{\ref{inst6},\ref{inst7}} \and David Barrado\inst{\ref{inst11}} \and Isabelle Boisse\inst{\ref{inst5}} 
           \and Aldo S. Bonomo\inst{\ref{inst9}} 
            \and David J. A. Brown\inst{\ref{inst1}} \and Giovanni Bruno\inst{\ref{inst5}} \and Javiera Rey Cerda\inst{\ref{inst8}}
             \and Bastien Courcol\inst{\ref{inst5}} \and Magali Deleuil\inst{\ref{inst5}} \and Rodrigo F. D\'iaz\inst{\ref{inst8}} 
              \and Amanda P. Doyle\inst{\ref{inst1}} \and Guillaume H\'{e}brard\inst{\ref{inst3},\ref{inst4}}
               \and James Kirk\inst{\ref{inst1}} \and Kristine W. F. Lam\inst{\ref{inst1}} 
               \and Don L. Pollacco\inst{\ref{inst1}} \and Arvind Rajpurohit\inst{\ref{inst5}} \and Jessica Spake\inst{\ref{inst1}} 
               \and Simon R. Walker\inst{\ref{inst1}}
          }
          
   \institute{Department of Physics, University of Warwick,
              Gibbet Hill Road, Coventry, CV4 7AL, UK\\
              \email{d.j.armstrong@warwick.ac.uk}\label{inst1}
            \and
            Aix Marseille Universit\'e, CNRS, LAM (Laboratoire d'Astrophysique de Marseille) UMR 7326, 13388, Marseille, France \label{inst5}
            \and
            Instituto de Astrof\'isica e Ci\^{e}ncias do Espa\c co, Universidade do Porto, CAUP, Rua das Estrelas, PT4150-762 Porto, Portugal \label{inst2}
             \and
            Departamento de Astrof\'isica, Centro de Astrobiolog\'ia (CSIC-INTA), ESAC campus 28691 Villanueva de la Ca\~nada (Madrid), Spain \label{inst11}
         \and
        Univ. Grenoble Alpes, IPAG, F-38000 Grenoble, France \label{inst6}
           \and 
       CNRS, IPAG, F-38000 Grenoble, France \label{inst7}
             \and
       INAF -- Osservatorio Astrofisico di Torino, via Osservatorio 20, 10025 Pino Torinese, Italy \label{inst9}
       \and
       Observatoire Astronomique de l'Universit\'e de Gen\`eve, 51 chemin des Maillettes, 1290 Versoix, Switzerland \label{inst8}
            \and
     Institut d'Astrophysique de Paris, UMR7095 CNRS, Universit\'e Pierre \& Marie Curie, 98bis boulevard Arago, 75014 Paris, France  \label{inst3}
          \and 
        Observatoire de Haute-Provence, Universit\'e d'Aix-Marseille \& CNRS, 04870 Saint Michel l'Observatoire, France \label{inst4}
       \and
       CNRS, Canada-France-Hawaii Telescope Corporation, 65-1238 Mamalahoa Hwy., Kamuela, HI 96743, USA \label{inst10}
    }
    
       \date{Received ; accepted}

 
  \abstract
   {}
   {The K2 mission has recently begun to discover new and diverse planetary systems. In December 2014, Campaign 1 data from the mission was released, providing high-precision photometry for \mytilde22000 objects over an 80-day timespan. We searched these data with the aim of detecting more important new objects.}
   {Our search through two separate pipelines led to the independent discovery of K2-19b \& c, a two-planet system of Neptune-sized objects (4.2 and 7.2 $R_\oplus$), orbiting a K dwarf extremely close to the 3:2 mean motion resonance. The two planets each show transits, sometimes simultaneously owing to their proximity to resonance and the alignment of conjunctions.}
   {We obtained further ground-based photometry of the larger planet with the NITES telescope, demonstrating the presence of large transit timing variations (TTVs), and used the observed TTVs to place mass constraints on the transiting objects under the hypothesis that the objects are near but not in resonance. We then statistically validated the planets through the \texttt{PASTIS} tool, independently of the TTV analysis.}
   {}


   \maketitle
%

\section{Introduction}
With the steady release of data from the K2 satellite, several projects have begun to search for previously undiscovered planetary systems. A number of interesting systems have already come to light \citep{Crossfield:2015vm,Vanderburg:2014wi,ForemanMackey:2015vi}. For these systems we now have photometry that approaches the precision of the \emph{Kepler} prime mission, and crucially, of host stars much brighter than the typical \emph{Kepler} case. This promises the use of radial velocity and other techniques to add to our knowledge of these already interesting objects. In this work we present a two-planet system observed in K2 field 1. This system, K2-19 (EPIC201505350, RA: 11:39:50.476, DEC: +00:36:12.87, Kepmag 12.8), lies exceptionally close to the 3:2 mean motion resonance (MMR), so it has the potential to show particularly large TTVs (a concept first suggested by \citet{Agol:2005fp,Holman:2005jf} for the general case). In terms of period ratio, only one object is as yet known to be closer to this resonance (and does not show TTVs, due to a large libration period). K2-19 was originally presented as a candidate planetary system in \citet{ForemanMackey:2015vi} and is validated here using further observations.

The 3:2 MMR is especially significant in both solar system and extrasolar planetary systems.  For decades, Pluto and Neptune were classified as the only solar system resonant planet pair, and their orbits evolve inside of a 3:2 MMR.  The Grand Tack model, a scenario proposed to explain the current architecture of the inner solar system, asserts that Jupiter and Saturn were once captured into a 3:2 MMR while embedded in the nascent solar nebula \citep{waletal2011,pieetal2014}.  Furthermore, the first extrasolar planetary system ever confirmed, around PSR 1257+12 \citep{wolfra1992,wolszczan1994}, includes two planets whose orbits are tightly coupled and are very close to residing within the 3:2 MMR \citep{maletal1992,gozetal2005,caletal2006}.

Transiting exoplanets orbiting main sequence stars represent the majority of known planets, but usually lack the necessary constraints to allow one to definitively assign membership to an individual MMR\footnote{With limited constraints, one can more easily exclude systems from existing within MMRs \citep{verfor2012}.}.  A commonly-used definition of MMR between two planets is the resulting dynamical state when a particular linear combination of mean longitudes, longitudes of pericentre, and sometimes longitudes of ascending node librate (or oscillate) about $0^{\circ}$ or $180^{\circ}$ over a given time interval \citep[e.g.][]{murder1999}.  For K2-19 and the 3:2 MMR, this linear combination is represented by either the angle $\theta_1$ or $\theta_2$, where

\begin{eqnarray}
\theta_1 &=& 3 \lambda_{\rm c} - 2 \lambda_{\rm b} - \varpi_{\rm b}
\\
\theta_2 &=& 3 \lambda_{\rm c} - 2 \lambda_{\rm b} - \varpi_{\rm c}
.
\end{eqnarray}

\noindent with $\lambda$ the mean longitude and $\varpi$ the longitude of pericentre. Here, and throughout the paper, we label the inner planet as b and the outer as c. This widely-adopted definition has several shortcomings, which are outlined in Section 1 of \cite{petetal2013}.  A definition which overcomes these shortcomings is: the term {\it resonant} is a trajectory that evolves in the region of phase space surrounded by the separatricies of a given integrable system Hamiltonian \citep[see, e.g.][]{morby2002}.

Because time series of planetary mean longitudes and longitudes of pericentre are typically not available in extrasolar systems, a common practice is to use orbital period ratios by themselves as a proxy for resonance.  High frequencies of systems with ratios of 1.5 and 2.0 \citep{lisetal2011} are suggestive that the 3:2 and 2:1 commensurabilities represent a significant tracer of formation, regardless of whether those planetary candidates are actually locked inside of a MMR. Recently, the frequency of systems just outside of the 3:2 period commensurability has exhibited a distinct asymmetry \citep[e.g.][]{fabetal2014}, which has led to substantial theoretical scrutiny \citep{batmor2013,leeetal2013,petetal2013,chafor2014,deletal2014,dellas2014}.  The period ratio of K2-19b \& c as displayed in the K2 data is $1.503514^{+0.000052}_{-0.000057}$, among the closest systems to a 3:2 commensurability so far detected. 

We searched the Exoplanet Orbit Database \citep{Han:2014hn} for other systems close to this commensurability. The only system we could find with a closer normalised distance to resonance \citep[defined in][]{Lithwick:2012ud}, $\triangle$, was the Kepler-372cd pair \citep{Rowe:2014wz}, where $\triangle$ is \mytilde 0.0003, as compared to K2-19 with \mytilde 0.0023. However, neither Kepler-372c nor d exhibit TTVs during the \emph{Kepler} observations due to a particularly long predicted TTV libration period, \mytilde 70 years. Also worth noting is the Kepler-342cd pair \citep{Rowe:2014wz}, with a $\triangle$ of \mytilde 0.0027, which also does not show TTVs due to a longer libration period.

Interest in these special period ratios is motivated by both the possibility of making deductions about a system's formation channel and its long-term future stability.  Convergent migration in protoplanetary discs is an effective and popular MMR formation mechanism \citep{sneetal2001}, and can also achieve three-body resonances \citep{pealee2002,libtsi2011}, although some MMRs are harder to lock into than others \citep{reietal2012,tadetal2015}.  Forming the 3:2 MMR in particular through energy dissipation has been widely investigated \citep{papszu2005,hadvoy2010,emel2012,ogikob2013,wanji2014,zhaetal2014}.  Capture into MMRs through gravitational scattering alone -- after the dissipation of the protoplanetary disc -- occurs relatively rarely \citep{rayetal2008}.

In this work we characterise this multi-planet system, validate its planetary nature using observed TTVs and the \texttt{PASTIS} tool, and discuss the implications these have for future observations.

\section{Observations}
\subsection{K2}
\label{sectK2obs}
Observations were made with the \emph{Kepler} satellite as part of the K2 mission between BJD 2456811.57 and 2456890.33, spanning \mytilde 80 days. The K2 mission \citep{Howell:2014ju} is the survey now being conducted with the repurposed \emph{Kepler} space telescope, and became fully operational in June 2014. It is surveying a series of fields near the ecliptic, returning continuous high-precision data over an 80 day period for each field. Despite the reaction wheel losses that ended the \emph{Kepler} prime mission, K2 has been estimated to be capable of 80ppm precision for V=12 stars, close to the sensitivity of the primary mission. All data will be public, although at the time of writing only campaigns 0 and 1 have been released. Targets are provided by the Ecliptic Plane Input Catalogue (EPIC) which is hosted at the Mikulski Archive for Space Telescopes (MAST)\footnote{https://archive.stsci.edu/k2/} along with the available data products. 

Targets in K2 often display significant pointing drift over the K2 observations, typically on a timescale of 6 hours on which the spacecraft thrusters are fired. This leads to a major source of systematic noise in the lightcurve \citep{Vanderburg:2014bi}, the removal of which is the key part of our detrending method. The full method is explained in \citet{Armstrong:2015bn}, but is summarised here for clarity. Initially a fixed aperture, of radius 4 pixels in this case and shape as described in \citet{Armstrong:2015bn}, was centred on the brightest target pixel. A raw light curve is extracted directly from this aperture, with background subtracted using the median out-of-aperture pixels. Row and Column centroid variations are found for the time series. At this point points associated with spacecraft thruster firings are removed, and the remaining points decorrelated from the centroid variations. We decorrelate from both row and column centroids simultaneously, as they are not statistically independent. This process can leave systematic or instrumental noise in place \citep[see][]{ForemanMackey:2015vi}. In this case this noise seems to be weak compared to the intrinsic stellar variability, which occurs with a magnitude of \mytilde 1\%.

This stellar variability is removed, along with any longer period systematics, through the application of an iteratively fitted polynomial. A 3D polynomial is fit to successive 2 day wide regions, with the fit repeated for 20 iterations clipping points greater than $3\sigma$ from the best fit line at each iteration. This fit is then used to detrend a 5 hour region at its centre, and the process repeated for each 5 hour region. As the principal components of the instrumental noise show variations on order 10+ days in campaign 1 \citep{ForemanMackey:2015vi} they should be removed through this process, and crucially will not affect the transits. We found that for this target the best results were obtained by performing this polynomial flattening immediately after extracting the lightcurve, \emph{before} decorrelating the flux from the centroid motion. The resulting light curve is shown in Figure \ref{figlc}. Note that some transits of each planet occur simultaneously with the other through the dataset. The final such simultaneous transit is of increased duration, and its shape begins to show signs of both planets individually.

\subsection{NITES}
\label{sectNITESobs}

Due to its proximity to MMR, K2-19 had the potential to show significant TTVs (see Section \ref{sectTTVs}). As such we scheduled it for further ground based transit observations. The Near Infra-red Transiting ExoplanetS (NITES) Telescope is a semi-robotic $0.4$-m (f/10) Meade LX200GPS Schmidt-Cassegrain telescope installed at the ORM, La Palma. The telescope is mounted with a Finger Lakes Instrumentation Proline 4710 camera, containing a $1024\times1024$ pixels deep-depleted CCD made by e2v. The telescope has a FOV and pixel scale of $11\times11$ arcmin squared and $0.66\arcsec$ pixel$^{-1}$, respectively and a peak QE$>90\%$ at $800$ nm. For more details on the NITES Telescope we refer the reader to \citet{McCormac14}.

\begin{figure}
\resizebox{\hsize}{!}{\includegraphics{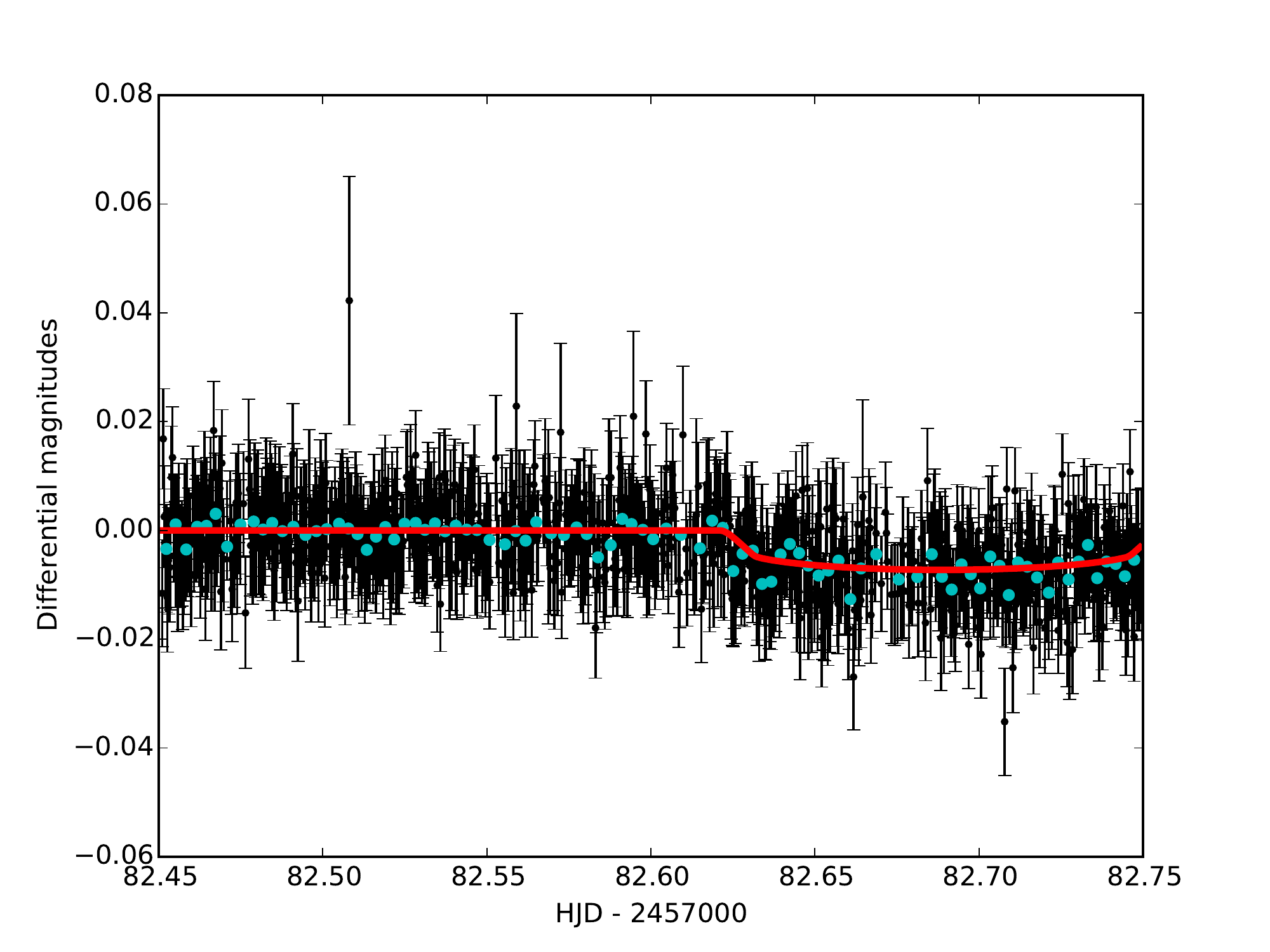}}
\caption{Data taken from the NITES telescope. The best fit transit derived from K2 observations is also plotted, along with the binned light curve, and fits the shape seen with NITES well.}
\label{figNITES}
\end{figure}

One transit of K2-19b was observed on 2015 Feb 28. The telescope was defocused slightly to $3.3\arcsec$ FWHM and $814$ images of $20$ s exposure time were obtained with $5$ s dead time between each. Observations were obtained without a filter. The data were bias subtracted and flat field corrected using PyRAF\footnotemark \footnotetext{PyRAF is a product of the Space Telescope Science Institute, which is operated by AURA for NASA.} and the standard routines in IRAF,\footnotemark \footnotetext{IRAF is distributed by the National Optical Astronomy Observatories, which are operated by the Association of Universities for Research in Astronomy, Inc., under cooperative agreement with the National Science Foundation.} and aperture photometry was performed using DAOPHOT \citep{1987PASP...99..191S}.  Ten nearby comparison stars were used and an aperture radius of $6.6\arcsec$ was chosen as it returned the minimum root mean square (RMS) scatter in the out of transit data. Initial photometric error estimates were calculated using the electron noise from the target and the sky and the read noise within the aperture. The data were normalised with a first order polynomial fitted to the out of transit data. The resulting lightcurve is shown in Figure \ref{figNITES}.

\subsection{SOPHIE}
\label{sectSOPHIE}
We observed the star K2-19 with the SOPHIE spectrograph mounted on the 1.93 m telescope at the Haute-Provence Observatory (France). We used the high-efficiency mode which has a spectral resolution of about 39 000 at 550nm. This mode is preferred for the observations of relatively faint stars \citep[e.g.][]{2014A&A...571A..37S}. For more information about the SOPHIE spectrograph, we refer the interested readers to \citet{2008SPIE.7014E..0JP} and \citet{2009A&A...505..853B}. We secured five epochs between 2015-01-23 and 2015-02-02 as part of our on-going TRANSIT consortium\footnote{OHP program: 14B.PNP.HEBR}. The exposure time ranges between 400s and 2700s, which lead to signal-to-noise ratio per pixel in the continuum at 550nm ranging between 13.5 and 22.1.

We cross correlated the spectra with a numerical mask corresponding of a G2 dwarf \citep{1996A&AS..119..373B,2002A&A...388..632P} and find a unique line profile with a width compatible with the rotational period found in K2 photometry (see next section). The derived radial velocity has a median uncertainty of 16ms$^{-1}$, and shows no significant variation.


\subsection{AstraLux}
\label{sectAstraLux}
In order to search for neighbouring stars which could be producing the transit signals, we obtain a high-spatial resolution image of K2-19 by applying the lucky-imaging technique with the AstraLux instrument installed at the 2.2m telescope at the Calar Alto Observatory (Almer\'ia, Spain). We obtained 65000 frames with an individual exposure time of 80 milliseconds in the SDSSz band. The images were reduced by using the observatory pipeline \citep[see][]{2008JPhCS.131a2051H}, which performs a basic reduction and selects the 10\% of frames showing the largest Strehl ratios \citep{1902AN....158...89S}. These frames are then aligned and stacked to produce the final high-spatial resolution image. The sensitivity curve of this image was obtained by adding artificial sources with different magnitude contrasts at different angular separations and measuring the recovery rate \citep[see][for further details]{Lillo-boxHRI2014}. We reach contrast magnitudes of $\Delta m_{z^{\prime}} = 5.5$ mag at 1\arcsec\, and $\Delta m_{z^{\prime}} = 6.3$ mag for angular separations larger than 1.5\arcsec\, at the $5\sigma$ level. No sources were found closer than 6\arcsec\, within our sensitivity limits. The analysis of the \textit{Kepler} centroids also revealed no significant shifts during transits above the \textit{Kepler} pixel size ($\sim$ 4\arcsec).

\section{Stellar parameters}

We obtained the parameters of the host star from the spectral analysis of five co-added
SOPHIE spectra. First, we subtracted from the spectra pointing to the source (in fibre A), 
any sky contamination using the spectra of fibre B, after correcting for the relative efficiency 
of the two fibres. The final spectrum has a S/N of the order of 25 around 6070\AA{}.

To derive the atmospheric parameters, namely the effective temperature ($T_{\mathrm{eff}}$), surface gravity ($\log g$), 
metallicity ($[Fe/H]$), and microturbulence (v$_{\textrm{mic}}$), we followed the methodology described in \cite{tsantaki2013}.
This method relies on the measurement of the equivalent widths (EWs) of Fe\,{\sc i} and Fe\,{\sc ii} lines and by imposing 
excitation and ionisation equilibrium. The analysis was performed in local thermodynamic equilibrium (LTE) using a grid of 
\citep{kurucz1993} model atmospheres and the radiative transfer code MOOG \citep{sneden1973}. 
Due to the low S/N of our spectrum, the EWs were derived manually using the IRAF splot task. 

From the above analysis, we conclude that the host star is a slightly metal-rich K dwarf.
The derived parameters are shown in Table \ref{tabstellarparams}. The stellar radius and mass were derived from the calibration of \citet{Torres:2010eoa}, updated with the version from \citet{Santos:2013dm} and the atmospheric parameters described above.
We also included in Table \ref{tabstellarparams} the determination of surface gravity from the transit fit parameters (see Section \ref{sectlcfit}) and the respective results of stellar mass and radius. We proceed using the transit derived parameters, as
they are much more accurate. 

We also study the stellar variability inherent in the lightcurve of K2-19. This lightcurve may be contaminated by remnant instrumental noise, but we find that repeating patterns apparent across the entirety of the K2 observations do not generally match the principal noise components seen \citep[see][for these components]{ForemanMackey:2015vi}. A weighted, floating mean Lomb-Scargle (LS) periodogram \citep{Lomb:1976bo,Scargle:1982eu}, following the method of \citet{Press:1989hb}, gives a principal period of 20.3 days (with an LS statistic of 323000, significantly above the background). Assuming this peak is due to stellar rotation then we are able to derive $P_{\textrm{rot}}$ as shown in Table \ref{tabstellarparams}. Errors on $P_{\textrm{rot}}$ are derived from the FWHM of the periodogram peak.

\begin{table}
\caption{Stellar parameters for K2-19}             
\label{tabstellarparams}      
\centering                         
\begin{tabular}{l l r}        
\hline\hline                 
Parameter & Value & Units \\    
\hline
$T_{\textrm{eff}}$ & $5230 \pm 417$ & K\\
log $g_{\star}$ & $4.39 \pm 0.79$ & dex \\
$v_{\textrm{mic}}$ & $0.92 \pm 0.5$ &  kms$^{-1}$\\
$[Fe/H]$ & $0.38 \pm 0.23$ & dex\\
$P_{\textrm{rot}}$ & $20.3^{+3.7}_{-2.3}$& days\\[5pt]
\hline
Derived Parameters: & & \\
$R_{\star}$ (spectroscopic) & $1.03 \pm 0.2$ & $R_\odot$ \\
$M_{\star}$ (spectroscopic) & $0.92 \pm 0.14$ & $M_\odot$ \\
log $g_{\star}$ (transit) & $4.52 \pm 0.22$ & dex \\
$R_{\star}$ (transit) & $0.88 \pm 0.06$ & $R_\odot$ \\
$M_{\star}$ (transit) & $0.89 \pm 0.06$ & $M_\odot$ \\
\hline                                  
\end{tabular}
\end{table}


\section{Light curve fitting}
\label{sectlcfit}
\begin{figure*}
\resizebox{\hsize}{!}{\includegraphics{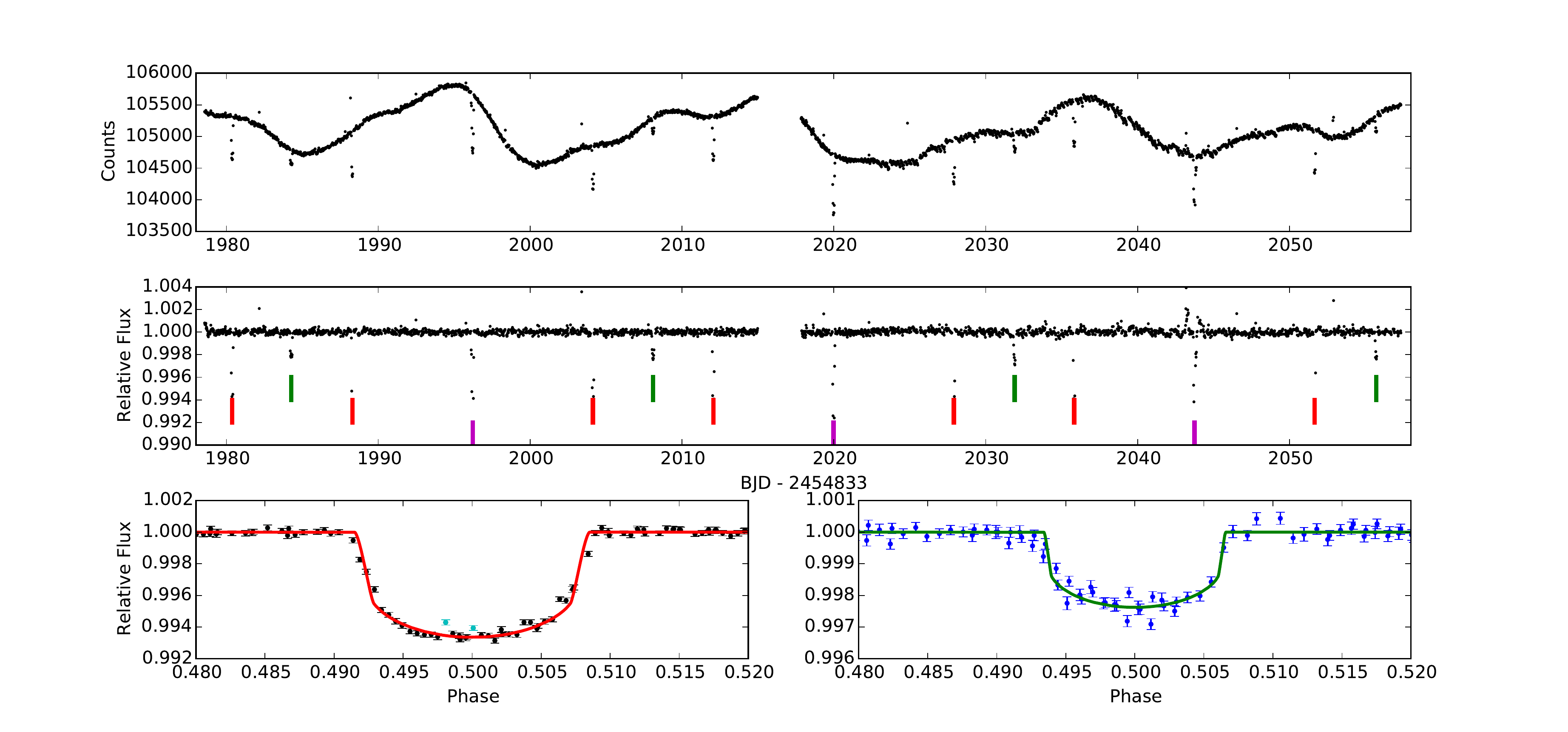}}
\caption{Top: Extracted light curve for K2-19, showing significant stellar variability. Middle: Flattened and detrended lightcurve, showing transits of the inner planet (b, red), outer planet (c, green) and simultaneous transits of both planets (magenta). Some outlier points are not shown for clarity. Bottom Left: The phase folded transits of planet b, excluding simultaneous transits and showing the best fit model. Some points (shown lighter than the others) displayed clear evidence of spot crossings by the planet and were excluded from the fit. Bottom Right: Same for planet c. Note the change in y-axis scale.}
\label{figlc}
\end{figure*}

To obtain the transit shape and parameters we limit ourselves to the K2 data, as it is of significantly higher precision and the NITES data do not show the full transit. The data were detrended as described in Section \ref{sectK2obs}, then cut so that only data within a 7 transit width region centred on each transit were used. We also removed all simultaneous transits, along with two specific points in separate transits which showed clear evidence for being within a spot crossing (significant brightenings within transit relative to their local transit shape). These points are highlighted in Figure \ref{figlc}. We note that there are other apparently bad points which were not removed - the decision to remove a point was based entirely on whether it was clearly anomalous within its local transit, to avoid excessive bias, and so points which only appear bad when shown phase folded and against the fit will remain. The data were then fit using the JKTEBOP code \citep[e.g.][]{Southworth:2013gc,Popper:1981de}, with numerical integration used to account for the long cadence of K2 observations (splitting each point into 60 integrated sub-points).

We initialised the fits with a linear limb darkening coefficient of 0.56, suitable for a K dwarf, which was then allowed to vary. We then tested for eccentricity, but found no constraint for either object. As such for the remaining tests the eccentricity of both planets was set to zero. To derive robust errors we used a Monte Carlo process whereby Gaussian observational errors are added to each data point and the fit repeated 1000 times, producing a distribution of best fits. The medians and 68.27\% confidence limits are then taken to produce values and errors. 

This process does not account for systematic errors in the light curve, such as starspots or correlated instrumental noise. These are of particular concern for K2-19; as has been noted there is evidence within some transits for spot crossings. In the past such crossings have proven useful in modelling starspots \citep[e.g.][]{2013MNRAS.430.3032B,Beky:2014df} but here they form a source of contamination to our fits. We test for the effect of these spots by adopting a prayer bead style residual permutation test. In this process, a best fit is acquired, and then the residuals of the data to this fit are `rolled' through the dataset, and a further best fit acquired each time. Due to the low cadence of K2 observations, there are not enough points near transit to get a distribution of parameter values through this method (270 and 183 tests respectively for planets b \& c). However, the prayer-bead generated distribution at least allows us to obtain an estimate of the systematic effect on our transit parameters. In all cases these systematic errors were comparable to or smaller than the Monte Carlo generated errors. As such we present final values and errors from the Monte Carlo tests. While we acknowledge that this method of testing for systematic errors is merely an estimate (especially as the full effect of spots only appears in transit), we note that as the errors generated by the prayer-bead process are not significantly larger that those from the Monte Carlo tests, the effect of systematics on the transit parameters is not particularly strong. 

The resulting best fits are shown for each planet in Figure \ref{figlc}. Note that the derived ephemeris are taken from only a small part of the TTV phase curve and so will require correction; see Section \ref{sectdiscuss} for detail. In particular, there are significantly larger errors on the period when TTVs are taken into account - final values are found in Section \ref{sectTTVanalysis}.

\begin{table*}
\caption{System parameters}             
\label{tabparams}      
\centering                         
\begin{tabular}{l l l r}        
\hline\hline                 
Parameter &  Units & b & c \\    
\hline                       
Model Parameters: & & &\\
P & days & $7.919454^{+0.000081}_{-0.000078}$ &  $11.90701^{+0.00039}_{-0.00044}$\\[5pt]
$T_0$ & BJD$_{TDB}-2456000 $& $813.38345^{+0.00036}_{-0.00039}$&  $817.2759 \pm 0.0012$\\[5pt]
$R_p/R_{\star}$ & & $0.0753^{+0.0028}_{-0.0015}$ & $0.0439^{+0.0011}_{-0.0012}$\\[5pt]
$(R_p+R_{\star})/a$ & & $0.0572^{+0.0084}_{-0.0042}$&  $0.0414^{+0.0015}_{-0.0009}$\\[5pt]
i & deg & $88.83^{+1.08}_{-0.89}$&  $89.91^{+0.05}_{-0.32}$\\[5pt]
e & & 0 (adopted) & 0 (adopted)\\[5pt]
Limb-Darkening & & 0$.552 \pm  0.041$&  $0.57^{+0.14}_{-0.13}$\\[5pt]

\hline
Derived Parameters: & & &\\
$R_p$ &$R_\oplus$ &$7.23^{+0.56}_{-0.51}$ & $4.21 \pm 0.31$\\[5pt]
a & AU & $0.077^{+0.008}_{-0.013}$ & $0.1032^{+0.0074}_{-0.0080}$\\[5pt]
 $S_{inc}$ & $S_\oplus$ & $87.7^{+ 9.3}_{-12.9}$ & $48.8^{+6.4}_{-6.2}$ \\[5pt]
 $P_c$/$P_b$ & & $1.503514^{+0.000052}_{-0.000057}$& \\[5pt]
 $\triangle$&  & $0.00234 \pm 0.00002$ & \\[5pt]
 
\hline                                  
\end{tabular}
\tablefoot{$\triangle$ is defined in Section \ref{sectdiscuss}, and represents the normalised distance to resonance. Note that $P_b$,$P_c$, and parameters derived from them are only instantaneous measurements, and will change over the course of the TTV phase curve (see Section \ref{sectdiscuss}). Transit based stellar parameters are used for derived quantities. }
\end{table*}

\section{Transit timing variations}
\label{sectTTVs}
Given the periods found in Section \ref{sectlcfit}, it is immediately apparent that the two planets in the K2-19 system lie close to the 3:2 MMR. It is common for systems close to MMR to show particularly large TTVs \citep{Lithwick:2012ud,Xie:2014jk} and so we searched the data in the hope of seeing variations. This search was carried out using the transit shape defined by our best fit parameters. As before, simultaneous transits were ignored. Points marked previously as being clear spot crossings were also excluded. We cut the data to a region within 2 full transit widths of the approximate transit centres, then passed the model transit over this region with a resolution of 0.00015 days. The minimum $\chi^2$ of this test series was recorded, at which point each datapoint was perturbed by a random Gaussian with standard deviation equal to the point error. The fit was then repeated, and this process undergone for 1000 iterations, to get a distribution of transit times. The mean of the distribution is then taken as the transit time. As when fitting the transit shape, this process does not account for systematic errors. This is particularly concerning for measuring transit times, because due to the low cadence of K2 observations only a few points are seen within each transit. If one of these points is significantly perturbed by a spot crossing (which occurs visibly for some transits) then the measured time would be strongly affected. To estimate the effect of these systematics, we repeat the prayer-bead residual analysis of Section \ref{sectlcfit}. In this case though, as we are considering each transit independently, there are even fewer data points near transit (typically \mytilde 30). Also of concern is that the full effect of spots can generally only be seen when they are occulted in transit, where there are even fewer points. As such we perform this analysis and estimate the systematic contribution to our error budget by taking the maximum and minimum parameter values which arise from the prayer bead test, over the \mytilde 30 iterations. We adopt these most pessimistic values as our $1\sigma$ errors, to ensure that we do not underestimate the errors on our transit times. The adopted values (from the mean of the monte carlo distribution) and errors (from the maximum and minimum of the prayer-bead residual test) are given in Table \ref{tabttimes}.

We were fortunate enough to obtain an additional transit of planet b with the NITES telescope (Section \ref{sectNITESobs}). The time of this transit was obtained using the transit shape derived from the K2 observations, meaning the only fit parameter was the time of transit centre. The same Monte Carlo test was performed, but now adopting the standard deviation of the resulting distribution as the error. We did not repeat the prayer-bead analysis in this case, as the transit did not show evidence for systematics.

\begin{figure}
\resizebox{\hsize}{!}{\includegraphics{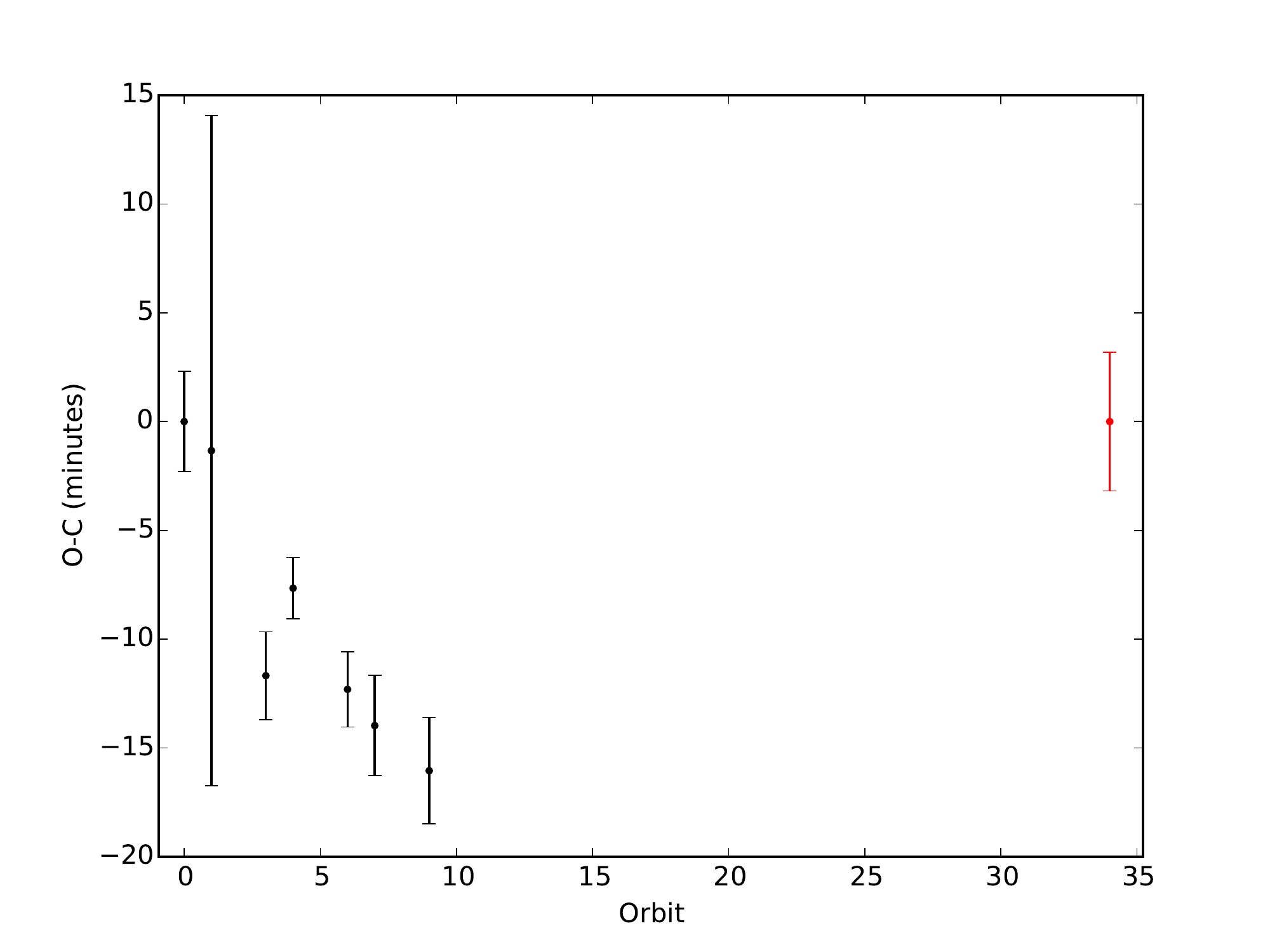}}
\caption{Observed-Calculated (O-C) transit times for planet b. Calculated times are taken by setting the O-C values at the first and last transits to zero because of the absence of well-determined period information. We do not use the \textit{K2}-derived ephemeris as these may misrepresent the scale of the TTVs.}
\label{figobj1TTV}
\end{figure}

\begin{figure}
\resizebox{\hsize}{!}{\includegraphics{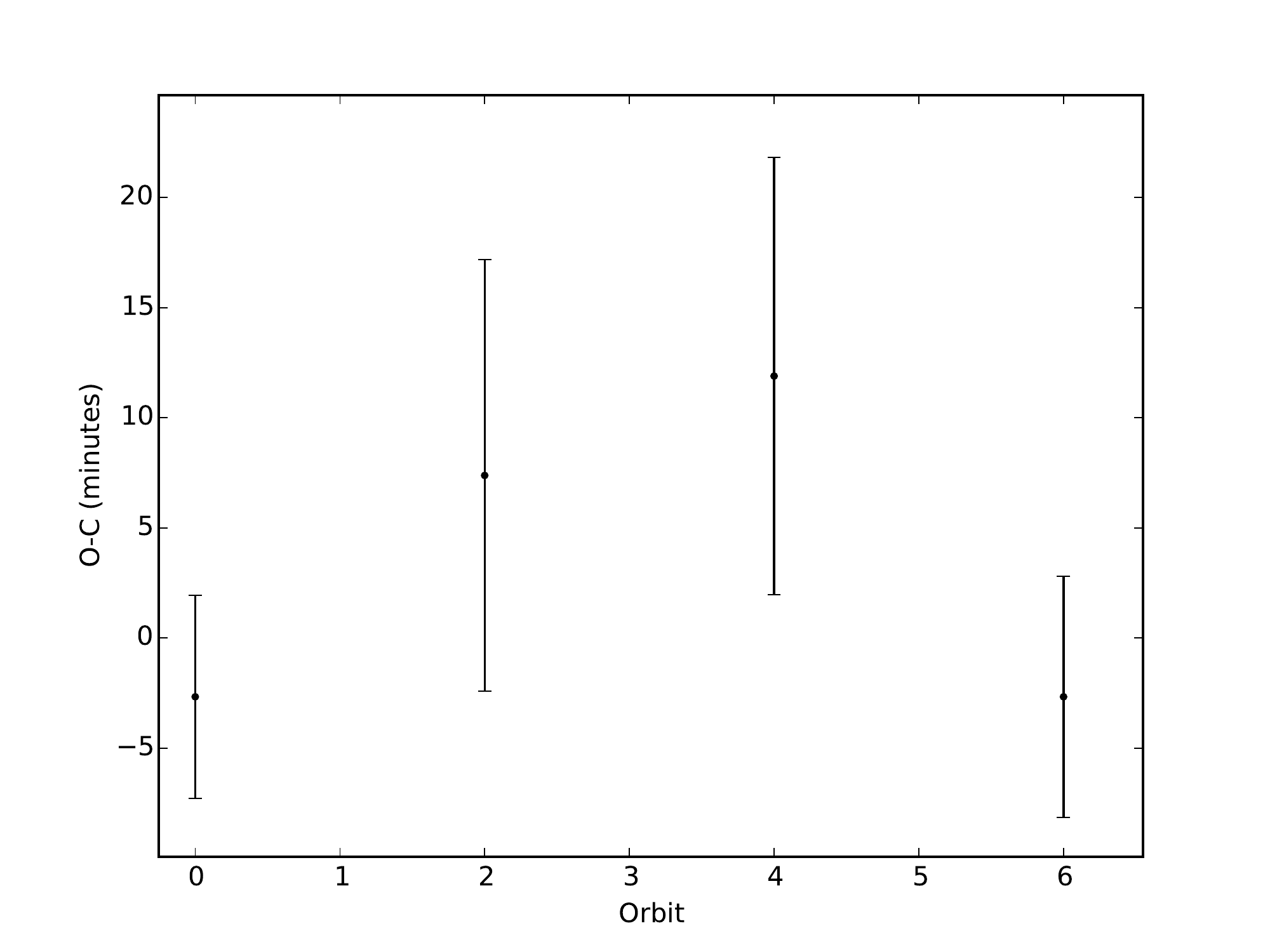}}
\caption{As Figure \ref{figobj1TTV} for planet c. In this case the calculated times are taken from the ephemeris of Table \ref{tabparams}.}
\label{figobj2TTV}
\end{figure}

The observed-calculated times found from our TTV analysis are shown in Figures \ref{figobj1TTV} and \ref{figobj2TTV}, calculated as described in the Figure captions. Planet b, in particular the NITES observation, show large TTVs of over an hour from the expected K2, linear ephemeris based, time (\mytilde a quarter of the transit duration). Within the K2 data alone we do not find the TTVs to be significant, beyond the third non-simultaneous transit for planet b which arrives earlier than would be expected. An initial analysis of these TTVs is performed in Section \ref{sectTTVanalysis}. We note that this detection of TTVs implies that the ephemeris in Table \ref{tabparams} \emph{are likely not the `true' ephemeris}, in the sense of the mean transit interval over long timeframes. Readers should thus be careful in predicting transit times. This is discussed further in Section \ref{sectdiscuss}.

\begin{table}
\caption{Detected transit times}             
\label{tabttimes}      
\centering                         
\begin{tabular}{l l l r}        
\hline\hline                 
Planet & Time (BJD$_{\textrm{TDB}}$-2456000) &  Error & Source \\    
\hline
b & 813.3841& 0.0016 & K2 \\
b & 821.3039& 0.0107 & K2 \\
b & 837.1382& 0.0014 & K2 \\
b & 845.06176& 0.00098 & K2 \\
b & 860.9000& 0.0012 & K2 \\
b & 868.8196& 0.0016 & K2 \\
b & 884.6597& 0.0017 & K2 \\
b &1082.6895 & 0.0022 & NITES\\
\hline
c & 817.2741 &0.0032& K2 \\
c & 841.0942 &0.0068& K2 \\
c & 864.9105 &0.0069& K2 \\
c & 888.7136 &0.0038& K2 \\
\hline                                  
\end{tabular}
\tablefoot{Simultaneous transits are not shown here.}
\end{table}

\section{\texttt{PASTIS} validation}
\label{sectplanetconf}

\subsection{Overview}
Candidates from \emph{Kepler} or \textit{K2} can be confirmed or validated in a number of ways. These include radial velocity observations, \texttt{BLENDER} and \texttt{PASTIS} analyses of the probability of false positive scenarios \citep{Torres:2010eob,2014MNRAS.441..983D}, high resolution photometry to search for close companions \citep[e.g.][]{Everett:2015fh,Lillo-boxMIT2012,Lillo-boxHRI2014,2014ApJ...791...35L} or TTVs \citep[e.g.][]{Steffen:2012kx,Nesvorny:2014ko}. 

Astrophysical false-positive scenarios such as eclipsing binaries might mimic the transit of a planet \citep{2005ApJ...619..558T}. Before claiming the planetary nature of a small and periodic signal, one should first rule out the possibility that this signal has a non-planetary origin. When two or more sets of transits are detected on the same target, their probability not to be planets significantly decreases \citep{Lissauer:2012hq, 2014ApJ...784...44L}. \citet{Rowe:2014wz} used this planet-likelihood ``multiplicity boost'' to validate a large sample of planets in multiple systems. This validation was possible because of the large sample of systems considered: a significant fraction of these systems are expected to be planets. \\

This argument cannot be used to validate a single multi-planetary system. Even if the two sets of transit in the K2-19 system are close to MMR, they could orbit two different stars. The period commensurability between the two could be the result of chance. For example, \citet{2014ApJ...784...44L} presented the interesting case of Kepler-132, a binary system in which each stellar member hosts a transiting planet at 6.2 and 6.4 days, respectively. That these two planets are orbiting with almost the same period is likely a coincidence. Therefore, the fact that the target star K2-19 hosts two planet-like objects close to MMR might also be a coincidence, so it is not evident that they are planets and that they both transit the same star, even if this is quite likely. In this section, we verify whether both transit sets have a planetary origin and orbit the target star.

\subsection{Definition of the scenarios and general framework}
\label{sectscenarios}
We identified 11 scenarios that could explain the \textit{K2} light curve, which we called $\mathcal{S}_{0}$ to $\mathcal{S}_{10}$:
\begin{itemize}
\item $\mathcal{S}_{0}$: the target star K2-19 is transited by two planets;
\item $\mathcal{S}_{1}$: a planet is transiting the target star K2-19 every 8 days and another planet is transiting a physical companion to the target star every 12 days;
\item $\mathcal{S}_{2}$: a planet is transiting the target star K2-19 every 12 days and another planet is transiting a physical companion to the target star every 8 days;
\item $\mathcal{S}_{3}$: two planets are transiting a physical companion to the target star;
\item $\mathcal{S}_{4}$: a planet is transiting the target star K2-19 every 8 days and an eclipsing binary with a period of 12 days is bound with the target star;
\item $\mathcal{S}_{5}$: a planet is transiting the target star K2-19 every 12 days and an eclipsing binary with a period of 8 days is bound with the target star;
\item $\mathcal{S}_{6}$: a planet is transiting the target star K2-19 every 8 days and an eclipsing binary with a period of 12 days is chance-aligned with the target star (background / foreground);
\item $\mathcal{S}_{7}$: a planet is transiting the target star K2-19 every 12 days and an eclipsing binary with a period of 8 days is chance-aligned with the target star (background / foreground);
\item $\mathcal{S}_{8}$: a planet is transiting the target star K2-19 every 8 days and another planet is transiting every 12 days a star chance-aligned with the target star (background / foreground);
\item $\mathcal{S}_{9}$: a planet is transiting the target star K2-19 every 12 days and another planet is transiting every 8 days a star chance-aligned with the target star (background / foreground);
\item $\mathcal{S}_{10}$: two planets are transiting a star chance-aligned with the target star (background / foreground).
\end{itemize}

To help the reader, a sketch of these scenarios is displayed in Figure \ref{Sketchs}. The scenario $\mathcal{S}_{0}$ is the one we want to test. The scenarios $\mathcal{S}_{1}$ to $\mathcal{S}_{5}$ invoke a stellar companion to the target star which is orbited either by planets or by eclipsing binaries. The scenarios $\mathcal{S}_{6}$ to $\mathcal{S}_{10}$ invoke a chance-aligned star transited by planets or eclipsed by lower-mass stars. For stability reasons, we did not consider here the case where two stars are eclipsing the same star at 8 and 12 days.\\

The probability of a scenario $\mathcal{S}_{i}$ is defined in the Bayesian framework as $\mathcal{P}\left(\mathcal{S}_{i}\, |\, \mathcal{D}, \mathcal{I}\right)$, with $\mathcal{P}$ the probability, $\mathcal{D}$ the data, and $\mathcal{I}$ the information. Using Bayes theorem, this probability can be expressed as:
\begin{equation}
\label{BayesTheorem}
\mathcal{P}\left(\mathcal{S}_{i}\, |\, \mathcal{D}, \mathcal{I}\right) = \frac{\mathcal{P}\left(\mathcal{S}_{i}\, |\, \mathcal{I}\right)\times\mathcal{P}\left(\mathcal{D}\, |\, \mathcal{S}_{i}, \mathcal{I}\right)}{\mathcal{P}\left(\mathcal{D}\, |\,\mathcal{I}\right)}.
\end{equation}
$\mathcal{P}\left(\mathcal{S}_{i}\, |\, \mathcal{I}\right)$ is the \textit{a priori} probability of the scenario $\mathcal{S}_{i}$, $\mathcal{P}\left(\mathcal{D}\, |\, \mathcal{S}_{i}, \mathcal{I}\right)$ is its marginalised posterior probability, and $\mathcal{P}\left(\mathcal{D}\, |\,\mathcal{I}\right)$ is a normalisation factor. The latter one being difficult to estimate, we instead computed the odds ratio $\mathcal{O}_{ij}$ between two scenarios $\mathcal{S}_{i}$ and $\mathcal{S}_{j}$, so that this normalisation factor is cancelled out:
\begin{equation}
\label{OddsRatio}
\mathcal{O}_{ij} = \frac{\mathcal{P}\left(\mathcal{S}_{i}\, |\, \mathcal{D}, \mathcal{I}\right)}{\mathcal{P}\left(\mathcal{S}_{j}\, |\, \mathcal{D}, \mathcal{I}\right)} = \frac{\mathcal{P}\left(\mathcal{S}_{i}\, |\, \mathcal{I}\right)\times\mathcal{P}\left(\mathcal{D}\, |\, \mathcal{S}_{i}, \mathcal{I}\right)}{\mathcal{P}\left(\mathcal{S}_{j}\, |\, \mathcal{I}\right)\times\mathcal{P}\left(\mathcal{D}\, |\, \mathcal{S}_{j}, \mathcal{I}\right)}.
\end{equation}
Then, assuming that $\sum_{i=0}^{10}\mathcal{P}\left(\mathcal{S}_{i}\, |\, \mathcal{D}, \mathcal{I}\right) = 1$, the probability of a given scenario $\mathcal{S}_{i}$ can be computed as:
\begin{equation}
\label{OddsSum}
\mathcal{P}\left(\mathcal{S}_{i}\, |\, \mathcal{D}, \mathcal{I}\right) = \left(\sum_{j=0}^{10}\mathcal{O}_{ji}\right)^{-1}.
\end{equation}
Therefore, to estimate the probability of the scenario $\mathcal{S}_{0}$, we need to compute the odds ratio between each pair of scenarios. For that, it is necessary to evaluate the \textit{a priori} probability of each scenario (see Section \ref{priorproba}) as well as their marginalised posterior probability (see Section \ref{posteriorproba}), which is defined as:
\begin{equation}
\mathcal{P}\left(\mathcal{D}\, |\, \mathcal{S}_{i}, \mathcal{I}\right) = \int\mathcal{P}\left(\vec{\theta}\, |\, \mathcal{S}_{i}, \mathcal{I}\right)\times\mathcal{P}\left(\mathcal{D}\, |\, \vec{\theta}, \mathcal{S}_{i}, \mathcal{I}\right)d\vec{\theta},
\end{equation}
with $\vec{\theta}$ the parameter space needed to model the scenario $\mathcal{S}_{i}$. Note that $\mathcal{P}\left(\mathcal{D}\, |\, \vec{\theta}, \mathcal{S}_{i}, \mathcal{I}\right)$ is also called the data likelihood which we computed assuming the data points are independent and normally distributed around their mean value $\left\{x_{1}, ..., x_{n}\right\}$ with a width of $\left\{\sigma_{1}, ..., \sigma_{n}\right\}$, with $n$ the number of data points:
\begin{equation}
\label{Likelihood}
\mathcal{P}\left(\mathcal{D}\, |\, \vec{\theta}, \mathcal{S}_{i}, \mathcal{I}\right) = \prod_{k=1}^{n}\frac{1}{\sqrt{2\pi}\sigma_{k}}\exp\left(-\frac{1}{2}\left(\frac{x_{k} - m_{k}\left(\vec{\theta}\, |\, \mathcal{S}_{i}\right)}{\sigma_{k}}\right)^{2}\right),
\end{equation}
with $m_{k}\left(\vec{\theta}\, |\, \mathcal{S}_{i}\right)$, the value of the model describing the scenario $\mathcal{S}_{i}$ that corresponds to the data point $x_{k}$.

\subsection{\textit{A priori} probabilities}
\label{priorproba}

To compute the probability of each scenario using the equations \ref{BayesTheorem} to \ref{Likelihood}, we first need to determine their \textit{a priori} probability. In the literature, there is no estimation of what is the occurrence rate of having, e.g. a two-planet system (like $\mathcal{S}_{0}$) or a planet in a triple system (like $\mathcal{S}_{4}$ and $\mathcal{S}_{5}$). Thus, we can not rely on robust statistical analysis of these scenarios to use as priors for our validation. To estimate these probabilities, we assume that each element of the systems has an independent probability to occur, i.e. the probability of having a planet is independent from that of having a stellar companion. This assumption is obviously wrong since stellar multiplicity should affect the formation of planets \citep[e.g.][]{2014ApJ...791..111W}, unless the stars are far enough from each other so that the impact is small. However, we assume this discrepancy does not significantly change our results.\\

For the stellar multiplicity, we used the results from \citet{2010ApJS..190....1R} who reported an occurrence of binary star systems of 34\% and an occurrence of triple systems of 9\%. Then, we used the results from \citet{2011arXiv1109.2497M} who reported an occurrence of planets at the level of 75\% (for any planets with periods up to 10 years). Assuming independence between the occurrence rates, we estimated that the \textit{a priori} probability to have, e.g. a target orbited by two planets is $\mathcal{P}\left(\mathcal{S}_{0}\, |\, \mathcal{I}\right) = 0.75\times0.75 = 0.5625$, or the probability to have two planets orbiting two different stars of a binary system is $\mathcal{P}\left(\mathcal{S}_{1}\, |\, \mathcal{I}\right) = \mathcal{P}\left(\mathcal{S}_{2}\, |\, \mathcal{I}\right) = 0.75\times0.75\times0.34 = 0.19125$. We listed the \textit{a priori} probabilities in the Table \ref{apriori}.\\

\begin{table}[h]
\caption{\textit{A priori} probability for the various scenarios given at the start of Section \ref{sectscenarios}.}
\begin{center}
\begin{tabular}{lrr}
\hline
\hline
Probability & Equation & Value\\
\hline
$\mathcal{P}\left(\mathcal{S}_{0}\, |\, \mathcal{I}\right)$ & $0.75\times0.75 $&$ 0.5625$\\
$\mathcal{P}\left(\mathcal{S}_{1}\, |\, \mathcal{I}\right)$ & $0.75\times0.34\times0.75 $&$ 0.19125$\\
$\mathcal{P}\left(\mathcal{S}_{2}\, |\, \mathcal{I}\right)$ & $0.75\times0.34\times0.75 $&$ 0.19125$\\
$\mathcal{P}\left(\mathcal{S}_{3}\, |\, \mathcal{I}\right)$ & $0.34\times0.75\times0.75 $&$ 0.19125$\\
$\mathcal{P}\left(\mathcal{S}_{4}\, |\, \mathcal{I}\right)$ & $0.75\times0.09 $&$ 0.0675$\\
$\mathcal{P}\left(\mathcal{S}_{5}\, |\, \mathcal{I}\right)$ & $0.75\times0.09 $&$ 0.0675$\\
$\mathcal{P}\left(\mathcal{S}_{6}\, |\, \mathcal{I}\right)$ & $0.75\times(4.6\times10^{-5})\times0.34 $&$ 1.17\times10^{-5}$\\
$\mathcal{P}\left(\mathcal{S}_{7}\, |\, \mathcal{I}\right)$ & $0.75\times(2.6\times10^{-5})\times0.34 $&$ 6.57\times10^{-6}$\\
$\mathcal{P}\left(\mathcal{S}_{8}\, |\, \mathcal{I}\right)$ &  $0.75\times(4.6\times10^{-5})\times0.75 $&$ 2.58\times10^{-5}$\\
$\mathcal{P}\left(\mathcal{S}_{9}\, |\, \mathcal{I}\right)$ & $0.75\times(2.6\times10^{-5})\times0.75 $&$ 1.45\times10^{-5}$\\
$\mathcal{P}\left(\mathcal{S}_{10}\, |\, \mathcal{I}\right)$ & $(2.6\times10^{-5})\times0.75\times0.75 $&$ 1.45\times10^{-5}$\\
\hline
\hline
\end{tabular}
\end{center}
\label{apriori}
\end{table}%

\begin{figure}[h!]
\begin{center}
\resizebox{\hsize}{!}{\includegraphics{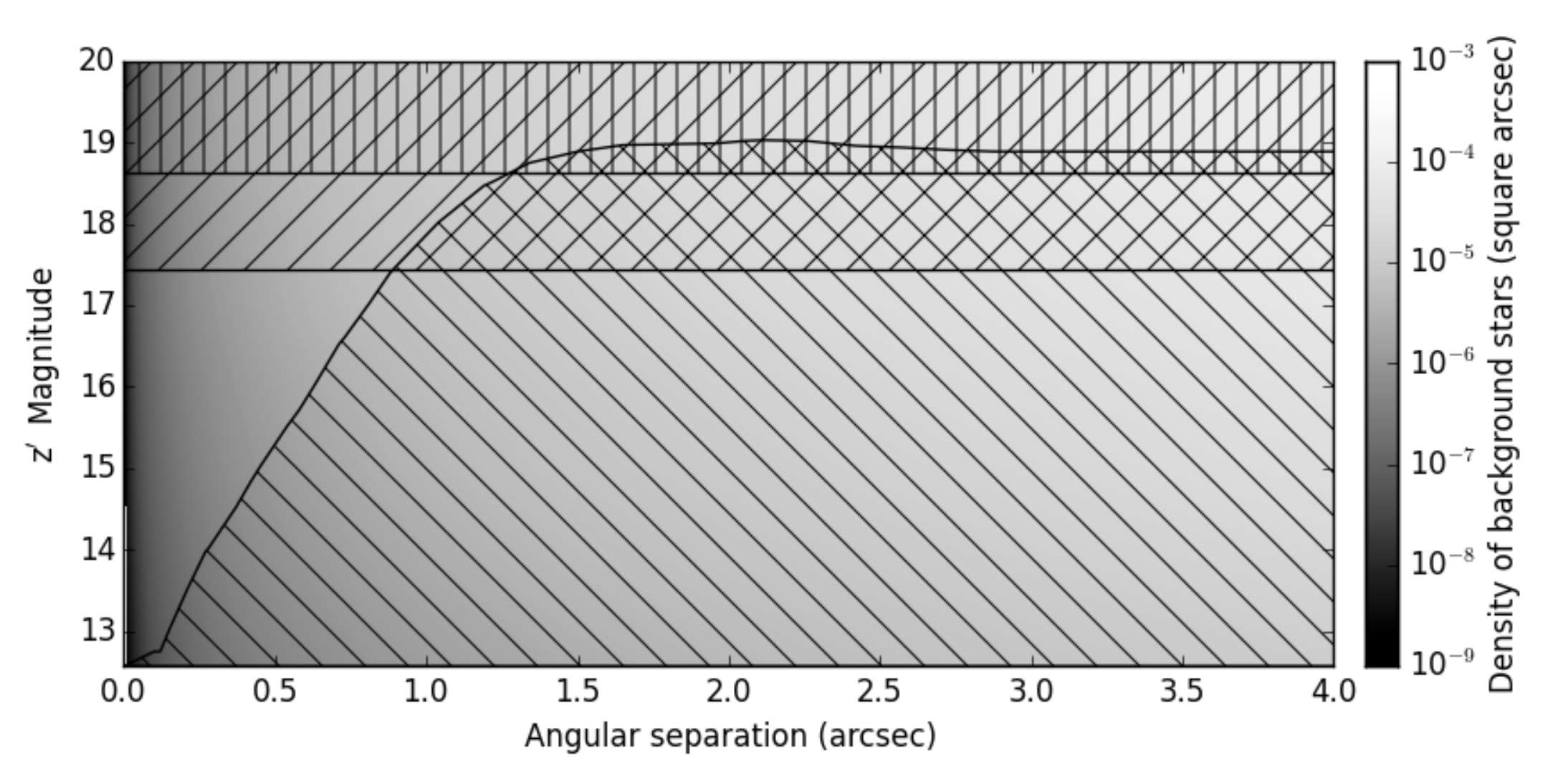}}\\
\caption{Map of the density of background stars chance-aligned with K2-19, integrated within an angular separation of up to 4\arcsec, as a function of its magnitude in the z$^{\prime}$-band. The negative-slope hatched region displays all the stars that would have been significantly detected in the AstraLux lucky-imaging data with more than 5-$\sigma$. The negative-slope and vertical hatched regions display the stars that are too faint to reproduce the observed transit depth of K2-19b \& c, respectively.}
\label{BGfig}
\end{center}
\end{figure}

To estimate the \textit{a priori} probability to have a background source of false positive (scenarios $\mathcal{S}_{6}$ to $\mathcal{S}_{10}$), we use the AstraLux data (Section \ref{sectAstraLux}). This data confirmed that no nearby star hosts one of the transiting objects. However, the data does not completely rule out the possibility that a background or foreground star, chance-aligned with the target, hosts one of the transits. To evaluate this probability, we used the Besan\c con galactic model \citep{2003A&A...409..523R} and simulated a stellar catalog of one square degree around the target star. We considered all types of stars with an apparent magnitude between 10 and 20 in the z$^{\prime}$ band. We set the interstellar extinction to zero magnitude per kpc in the simulation, which we corrected \textit{a posteriori} using the galactic extinction model of \citet{2005AJ....130..659A}. We derived the expected density of stars in the vicinity of K2-19 per bin of magnitudes within different angular separations from the target star. This density of stars is plotted in Figure \ref{BGfig}, together with the 5-$\sigma$ detection limits from AstraLux lucky imaging. 

A background source of false positives can only mimic the transit depths of the planets if its magnitude is within a range $\Delta m$ defined as  \citep{2011ApJ...738..170M}:
\begin{equation}
\Delta m = 2.5 \log_{10}\left(\frac{\delta_{tr}}{\delta_{bg}}\right),
\end{equation}
where $\delta_{tr}$ and $\delta_{bg}$ are the depths of the transit, as measured in the light curve, and the depth on the background star, respectively. Assuming a maximum eclipse depth of 50\% for the background star, we find that the maximum magnitude range is of 4.9 and 6.0 for the transit signal at 8 and 12 days (respectively). The target star is of magnitude z$^{\prime} = 12.6$, hence false positives can probe stars as faint as magnitude 17.4 and 18.6 (respectively) in the same bandpass. These maximum magnitude limits are also represented in the Fig. \ref{BGfig}. By integrating the expected number of stars that are bright enough to mimic the transit depths if they were eclipsing binaries within the detection limits of AstraLux, we find $2.6\times10^{-5}$ stars that might mimic the 8-d signal and $4.6\times10^{-5}$ stars that might mimic the 12-d signal. To derive the \textit{a priori} probability of the scenarios $\mathcal{S}_{6}$ to $\mathcal{S}_{10}$, we therefore multiply these values with the probability of being an eclipsing binary or that of hosting a planet. These values are reported in Table \ref{apriori}. Note that the \textit{a priori} probability for scenarios $\mathcal{S}_{8}$ to $\mathcal{S}_{10}$, where the chance-aligned star is transited by one or both planets is overestimated since no transiting planet has been found so far with a transit depth as large as 50\%.

\subsection{The posterior distribution}
\label{posteriorproba}

To estimate the posterior probability that the K2-19 data are produced by one of the aforementioned scenarios, we used the \texttt{PASTIS} software \citep{2014MNRAS.441..983D, 2015MNRAS.451.2337S} to model the \textit{K2} photometric measurements\footnote{We used only the data collected in the vicinity of non-simultaneous transits.}. The light curve was modelled using the \texttt{EBOP} code \citep{1972ApJ...174..617N, 1981psbs.conf..111E, Popper:1981de} extracted from the \texttt{JKTEBOP} package \citep{2008MNRAS.386.1644S}. For the limb darkening coefficient, we used the interpolated values from \citet{2011A&A...529A..75C}. To model the stars \citep[see][for a more detailed description on how we model the stars in \texttt{PASTIS}]{2014MNRAS.441..983D}, we used the Dartmouth stellar evolution tracks of \citet{2008ApJS..178...89D} and the BT-SETTL stellar atmosphere models of \citet{2012IAUS..282..235A} which we integrated in the \textit{Kepler} bandpass. We used the results from the spectroscopic analysis for the parameters of the target star and the initial mass function from \citet{2001MNRAS.322..231K} for the blended stars. The stars that are defined as gravitationally bound are assumed to have the same metallicity and the same age. The orbits are assumed to be circular. We assumed that the blended stars are at least fainter by 1 magnitude in the V-band than the target star, otherwise any such star would have been clearly identified in the spectroscopic data. The \textit{K2} light curve was modelled allowing the out-of-transit flux, contamination and an extra source of white noise (jitter) to vary. We used an oversampling factor of 10, to account for the finite integration time of the \textit{Kepler} long-cadence data \citep{2010MNRAS.408.1758K}. We also model the spectral energy distribution of K2-19 composed by the magnitudes in the Johnson B and V, Sloan g$^{\prime}$ and i$^{\prime}$, 2-MASS J, H, and Ks, and WISE W1 to W3 bandpasses from the APASS database \citep{2015AAS...22533616H} and the AllWISE catalog \citep{2010AJ....140.1868W}.\\

We analyse the aforementioned data using the MCMC procedure described in \citet{2014MNRAS.441..983D}. For the orbital ephemeris, we used Normal priors matching the ephemeris reported in Table \ref{tabparams} with uncertainties boosted by 100, to avoid biasing the results with too narrow priors. For the other parameters, we choose uninformative priors. We limit the priors on the planet radius to be less than 2.2 R$_{jup}$, which is the radius of the biggest planet found to date: KOI-13 \citep{2011ApJ...736L...4S}. The exhaustive list of parameters and their priors are reported in Table \ref{PASTISpriors}. For all scenarios, we ran a minimum of 10 chains of 1.10$^{6}$ iterations, randomly started from the joint prior distribution. We then selected the chains that converged toward the maximum likelihood, which we thinned, to derive the posterior distributions for the 11 scenarios. If the posterior distributions had less than 1000 independent samples, we ran new chains until reaching this threshold. For the scenarios $\mathcal{S}_{6}$ to $\mathcal{S}_{10}$, we needed to run up to 40 chains to reach the threshold. We report in Table \ref{PASTISresults} the median and 68.3\% confidence interval for the free parameters. All the fitted parameters for the scenario $\mathcal{S}_{0}$ are compatible within 1-$\sigma$ with those derived in Section \ref{sectlcfit}.\\

\subsection{Scenario probability and planet validation}

We marginalised the posterior distribution $\mathcal{P}\left(\vec{\theta}\, |\, \mathcal{S}_{i}, \mathcal{I}\right)\times\mathcal{P}\left(\mathcal{D}\, |\, \vec{\theta}, \mathcal{S}_{i}, \mathcal{I}\right)$ over the parameter space using the method described in \citet{2012A&A...544A.116T}. As discussed by the authors, this method underestimates the Occam's razor when computing the evidence. However, as already discussed in \citet{2014MNRAS.441..983D} and \citet{2014A&A...571A..37S}, this limitation is expected to be relatively weak here since most scenarios have the same number of free parameters and most parameters share the same priors (e.g. target parameters, orbital parameters). Therefore, we assumed our results are not dependent on the method used to marginalise the posterior distribution.\\

We then computed the probability of each scenario using Equation \ref{OddsSum}, which we multiplied by the \textit{a priori} probabilities listed in Table \ref{apriori}. These scenarios' probabilities are reported in Table \ref{probascenarios}. The scenario $\mathcal{S}_{0}$ (i.e. two planets transiting the target star) has a probability of 99.2\% while the other scenarios have a probability below 1\%. We can therefore conclude that (1) the two objects producing the transits seen in the K2-19 light curves are planets, (2) the two planets are transiting the same star, and (3) that star is the target star K2-19. As such we are able to statistically validate the two planets K2-19b \& c.

\begin{table}[h]
\caption{Probability for the various scenarios. Note that the probabilities of the scenarios $\mathcal{S}_{1}$ to $\mathcal{S}_{10}$ are so low that we provide instead their logarithmic values. The logarithmic value of scenario $\mathcal{S}_{0}$ is provided for comparison.}
\begin{center}
\begin{tabular}{cc}
\hline
\hline
Probability & Value\\
\hline
$\mathcal{P}\left(\mathcal{S}_{0}\, |\, \mathcal{D}, \mathcal{I}\right)$ & 99.2$^{_{+0.4}}_{^{-0.8}}$\%\\
$\log_{10}\left(\mathcal{P}\left(\mathcal{S}_{0}\, |\, \mathcal{D}, \mathcal{I}\right)\right)$ & -0.0035$^{_{+0.0017}}_{^{-0.0035}}$\\
$\log_{10}\left(\mathcal{P}\left(\mathcal{S}_{1}\, |\, \mathcal{D}, \mathcal{I}\right)\right)$ & -2.17 $\pm$ 0.30\\
$\log_{10}\left(\mathcal{P}\left(\mathcal{S}_{2}\, |\, \mathcal{D}, \mathcal{I}\right)\right)$ & -2.90 $\pm$ 0.32 \\
$\log_{10}\left(\mathcal{P}\left(\mathcal{S}_{3}\, |\, \mathcal{D}, \mathcal{I}\right)\right)$ & -3.71 $\pm$ 0.34 \\
$\log_{10}\left(\mathcal{P}\left(\mathcal{S}_{4}\, |\, \mathcal{D}, \mathcal{I}\right)\right)$ & -3.80 $\pm$ 0.31 \\
$\log_{10}\left(\mathcal{P}\left(\mathcal{S}_{5}\, |\, \mathcal{D}, \mathcal{I}\right)\right)$ & -4.42 $\pm$ 0.44\\
$\log_{10}\left(\mathcal{P}\left(\mathcal{S}_{6}\, |\, \mathcal{D}, \mathcal{I}\right)\right)$ & -5.03 $\pm$ 0.54 \\
$\log_{10}\left(\mathcal{P}\left(\mathcal{S}_{7}\, |\, \mathcal{D}, \mathcal{I}\right)\right)$ & -5.17 $\pm$ 0.38\\
$\log_{10}\left(\mathcal{P}\left(\mathcal{S}_{8}\, |\, \mathcal{D}, \mathcal{I}\right)\right)$ & -4.41 $\pm$ 0.41 \\
$\log_{10}\left(\mathcal{P}\left(\mathcal{S}_{9}\, |\, \mathcal{D}, \mathcal{I}\right)\right)$ & -3.87 $\pm$ 0.39\\
$\log_{10}\left(\mathcal{P}\left(\mathcal{S}_{10}\, |\, \mathcal{D}, \mathcal{I}\right)\right)$ & -6.00 $\pm$ 0.32\\
\hline
\hline
\end{tabular}
\end{center}
\label{probascenarios}
\end{table}%

\section{TTV mass limits and stability}
\label{sectdiscuss}
\subsection{Transit timing variations - overview}
\label{sectTTVoverview}
We leave a detailed study of the TTVs to future analysis (Barros et al in prep). It is however possible to place a number of constraints on the system even with the limited coverage of the TTV phase curve which we obtain here. For this initial analysis, we use the analytical representation of the TTVs derived by \citet{Lithwick:2012ud}, hereafter L12, which has been shown to be valid for systems near MMR \citep{Deck:2014wm}, a condition strongly met in this case. We note that the L12 formulae are only valid for objects not explicitly `in' resonance. Without further constraints it is impossible to tell whether this condition is met here. As such this analysis proceeds under the assumption that the objects are not in resonance. The use of the L12 formulae allows us to obtain a more intuitive description of the parameter space than is generally possible using N-body simulations. Given the potential for spots or other systematic errors to affect the \textit{K2} transit times, and the otherwise limited coverage of the TTV phase curve, we defer such an analysis to future work.

The TTV phase curve described by L12 is a sinusoid with two key parameters: (1) an amplitude $|V|$ given as a function of planetary mass, stellar mass, $\triangle$ (the normalised distance to resonance), and the free eccentricity $Z_{free}$ (a complex number), and (2) a period given by

\begin{equation} 
P_{\textrm{super}} = \frac{P_{\textrm{outer}}}{j|\triangle|}
\end{equation}

where 

\begin{equation}
\triangle = \frac{P_{\textrm{outer}}}{P_{\textrm{inner}}}\frac{j-1}{j} - 1
\end{equation}

For the 3:2 MMR $j=3$. This leads to a phase curve of the form

\begin{equation}
TTV = |V|\sin(\frac{t - t_0}{P_{\textrm{super}}} + \phi)
\end{equation}

\noindent where $\phi$ is the phase of the curve and changes over the secular timescale (hence is constant for our purposes). In our case, we can set $t_0$ to be the time of first transit to acceptable accuracy, due to the alignment of planetary conjunctions demonstrated by the simultaneous transits observed. Both the amplitude and period depend strongly on $\triangle$. The closer a system is to resonance, the larger the amplitude becomes, but the longer the period. For a system as close to resonance as K2-19, the period is particularly long, of the order several years. This means that within the 80 days of \textit{K2} observations, we would not expect to see large variation. With the later NITES transit however, we are starting to see the high amplitude TTV curve that these planets exhibit.

The period $P_{\textrm{super}}$ depends only on $\triangle$, and the period of the outer body. The TTV amplitude however generally shows a degeneracy between the free eccentricity $Z_{free}$ and planetary mass (L12). \citet{Hadden:2014bf} break this degeneracy statistically, but for individual objects it can be difficult to circumvent (although so-called synodic chopping signals can help, see \citet{Nesvorny:2014gm}). Furthermore with our limited observations the current phase of the TTV curve $\phi$ is unclear. In the low free eccentricity case ($|Z_{free}|<<|\triangle|$), $\phi$ must be zero (L12), however there is no guarantee that this is the case here.

\subsection{Transit timing variations - analysis}
\label{sectTTVanalysis}
We here explore the parameter space allowed by this analytical representation, given the transit times we have observed. The analysis is most strongly constrained by the NITES transit, as this transit allows much more coverage of $P_{\textrm{super}}$. The process is complicated by a correction term which must be added to our derived periods. As they are derived from a limited part of the TTV phase curve, they do not represent the overall `true' period, in the sense of the mean transit interval over long timescales. This makes determination of $P_{\textrm{super}}$ non-trivial. Small corrections to the periods can change $\triangle$ significantly, which has a strong effect on the TTV period and amplitude. We circumvent this problem by utilising the fact that our derived periods are in fact measurements of the \emph{gradient} of the TTV curve at the time of the \textit{K2} observations. The sensitivity to $\triangle$ exhibited by the TTV period and amplitude cancels out when calculating the gradient, allowing the correction to the periods to be made using only the initial periods. Using this, we can fit our transit times using the following process:
\begin{enumerate}
\item Take values for the input parameters, $M_b$, $M_c$, $Re(Z_{\textrm{free}})$, $Im(Z_{\textrm{free}})$, $\phi$
\item Calculate the correction to make to the derived periods
\item Use the corrected periods to find the true $\triangle$ and $P_{super}$
\item Use the input parameters and the newly corrected values to calculate the TTV amplitude
\item Compare the now fully defined TTV curve to the observations
\end{enumerate}

We begin with the case where $|Z_{free}|<<|\triangle|$, and the free eccentricity can be ignored. This reduces the input parameters to solely the two planetary masses, as $\phi$ must also be zero in this case (L12). We set $t_0$ to be zero at the time of first transit - this is accurate to within \mytilde 20 days, and given the long $P_{\textrm{super}}$ of several years, this does not need to be more accurate for this analysis.

\begin{figure}
\resizebox{\hsize}{!}{\includegraphics{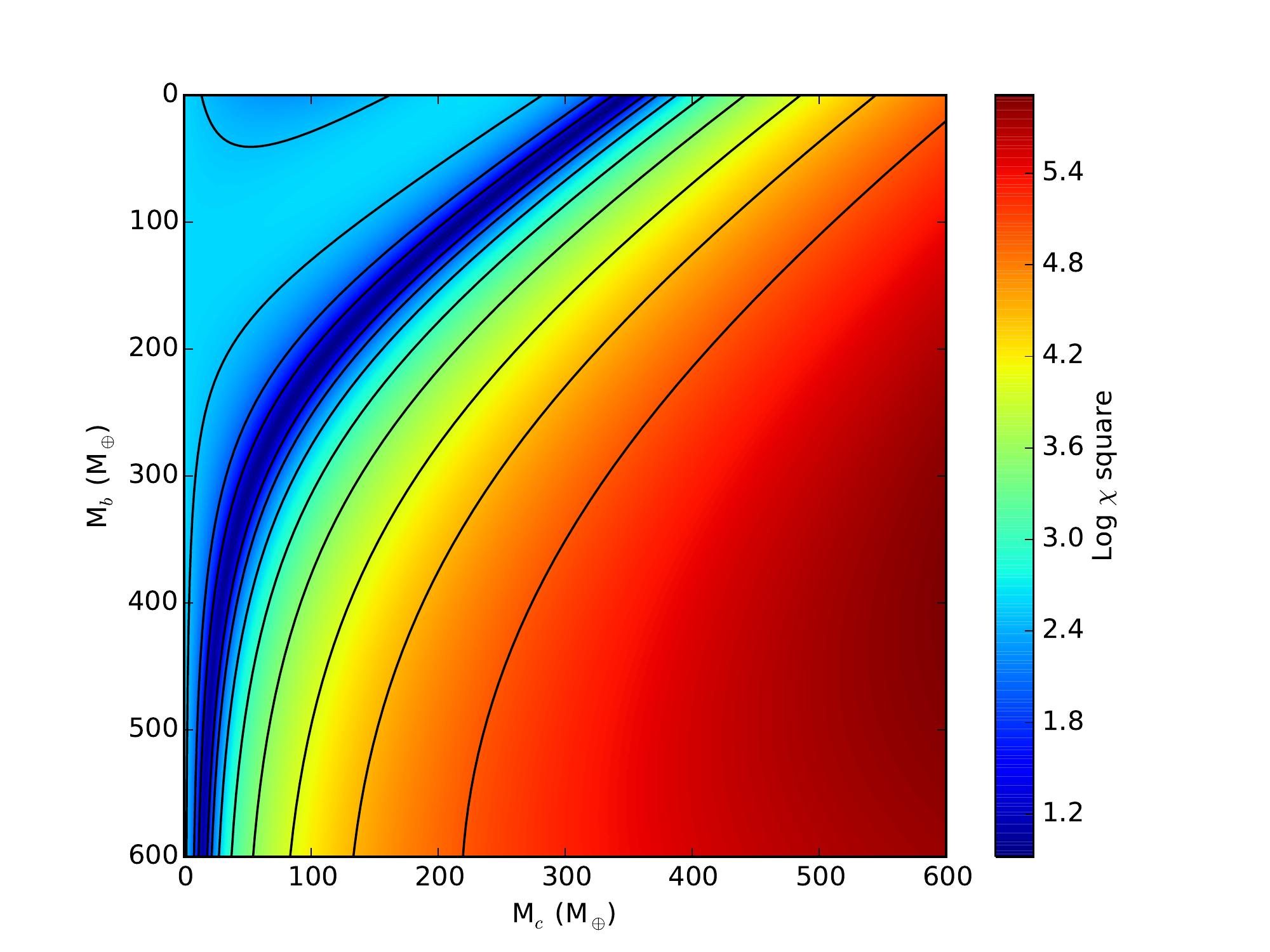}}
\caption{The $\chi^2$ surface given by the assumption that $|Z_{free}|<<|\triangle|$. A good fit is obtained for limited combinations of the two masses. Contour levels are shown at intervals of 0.5.}
\label{figchisqzfree0}
\end{figure}

\begin{figure}
\resizebox{\hsize}{!}{\includegraphics{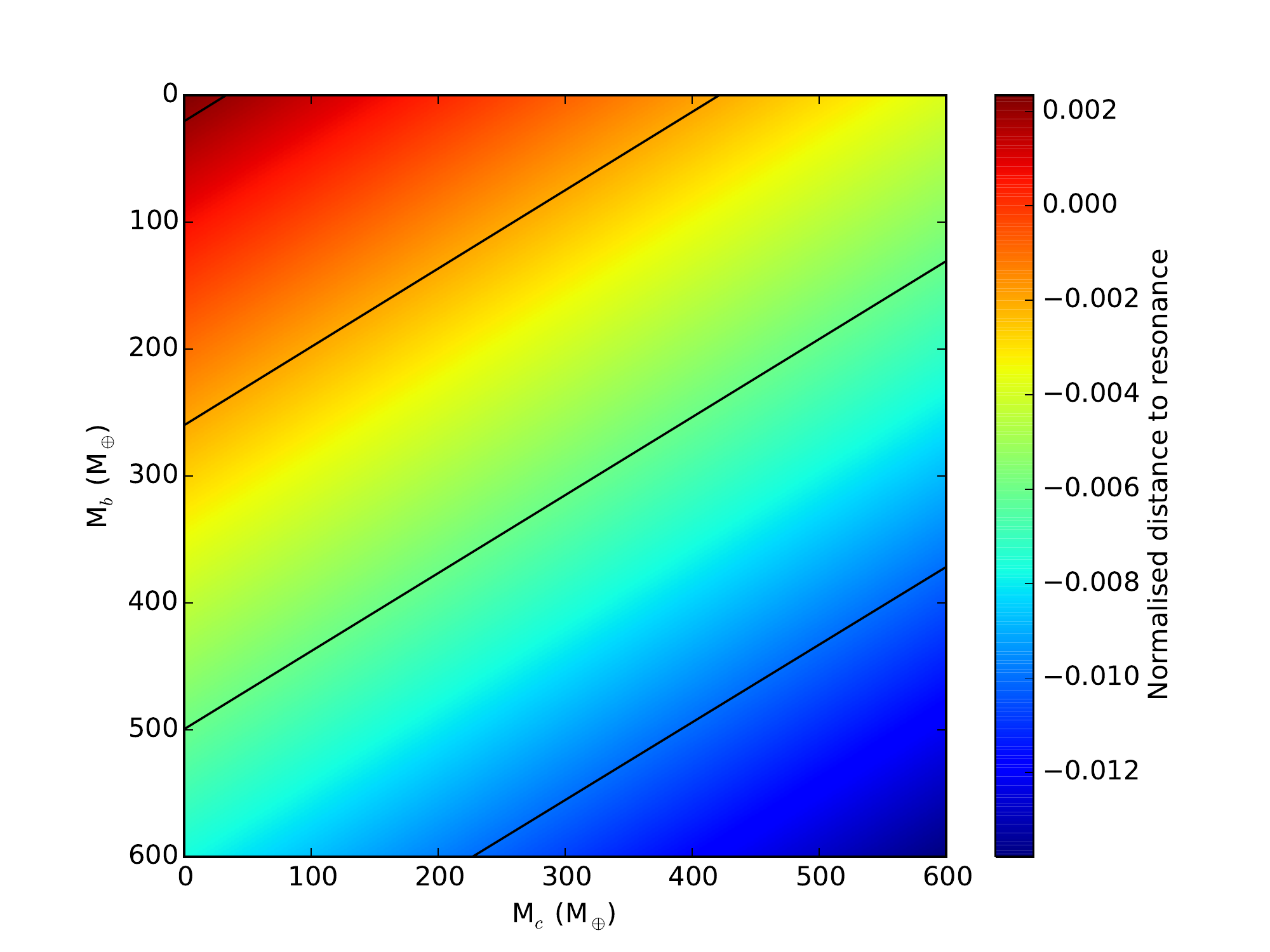}}
\caption{As Figure \ref{figchisqzfree0}, but showing the normalised distance to resonance $\triangle$ over the same mass ranges. In this case it can be seen that $\triangle$ becomes negative for all good fits, implying that the planets are in fact below the 3:2 resonance if this assumption held. Contour levels are shown at intervals of 0.004.}
\label{figdeltazfree0}
\end{figure}

The parameter space of the low eccentricity case is best shown in Figure \ref{figchisqzfree0}, which shows the log $\chi^2$ surface seen. The key points are a maximum mass for the outer planet c, of $350 M_\oplus$, found when the mass of the inner planet b goes to zero. The mass of planet b is less constrained (to constrain it properly using this analysis would require observations of planet c's TTV curve), but can only take high values in the event that planet c's mass drops much lower. Similarly $P_{\textrm{super}}$, excepting very high masses (over \mytilde $600M_\oplus$) for planet b, is constrained to be greater than \mytilde 480 days, and less than \mytilde 3050 days. The accompanying surface for the `true' $\triangle$ is shown in Figure \ref{figdeltazfree0}, and makes clear that all zero eccentricity best fitting planetary mass combinations imply a `true' $\triangle$ that is in fact slightly below the 3:2 resonance. As such while we cannot confirm this is the case from these observations alone, it is possible that the apparent planetary periods oscillate around the resonance over the course of the TTV curve. 

\begin{figure}
\resizebox{\hsize}{!}{\includegraphics{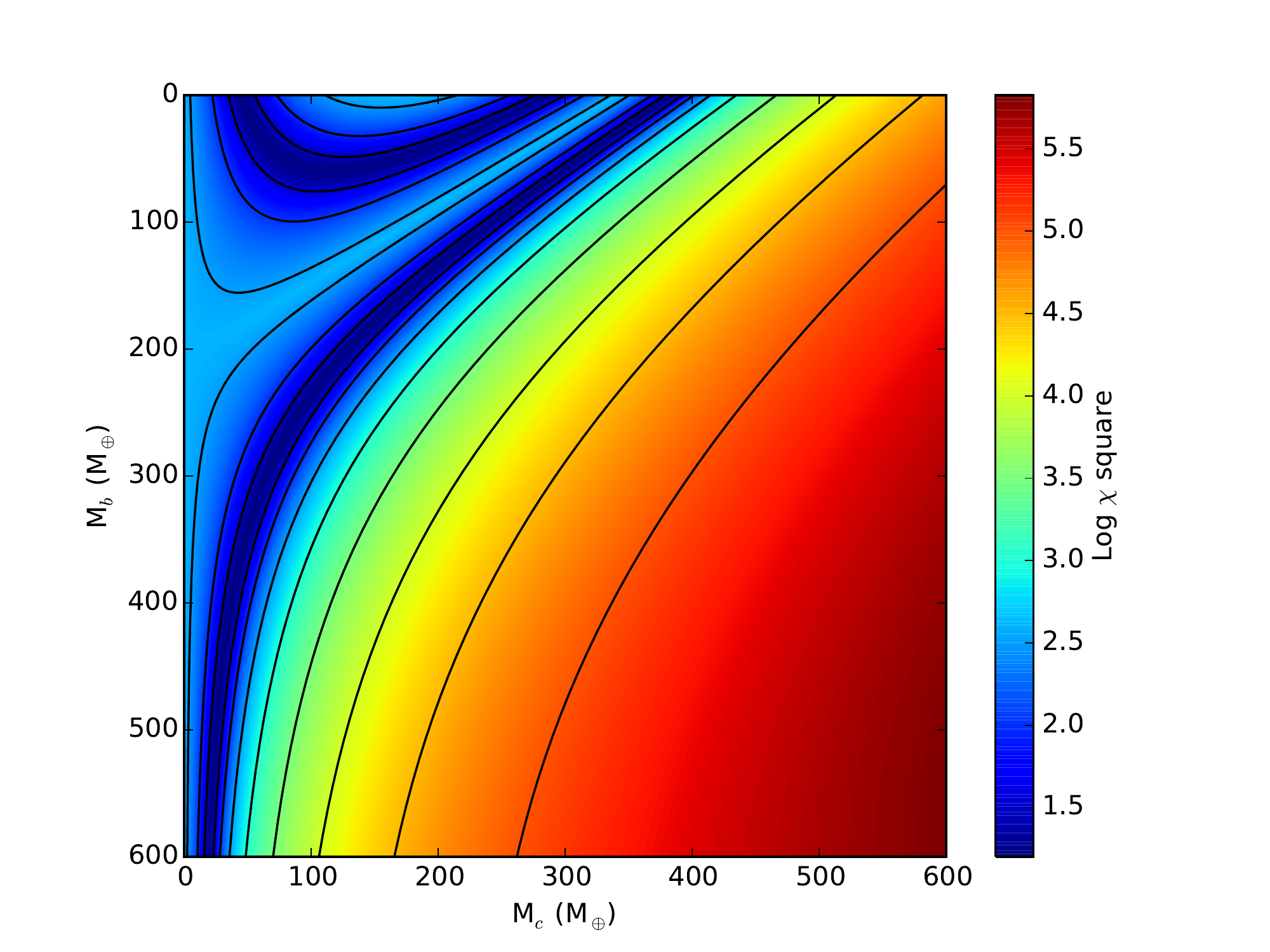}}
\caption{As Figure \ref{figchisqzfree0}, but showing the $\chi$ square surface for Re($Z_{\textrm{free}}$)=0.2, and $\phi=\pi/4$ over the same mass range. A degeneracy can be seen between positive $\triangle$ (top left) and negative $\triangle$ (down and towards the right). Contour levels are shown at intervals of 0.5.}
\label{figchisqphipiov4}
\end{figure}

Extending our analysis to the case where there is significant free eccentricity, we can immediately constrain $\phi$. Because the NITES transit arrived late rather than early, $\phi$ must be in the range $0 \le \phi \le \pi$. Within this range however a number of different effects can occur. We test these cases by repeating the analysis with the real component of $Z_{\textrm{free}}$ set to be 0.2, at various values of $\phi$. We hold the imaginary component at zero. Although the imaginary component can affect the amplitude and phase of the TTV curve, as we are trialling different values of $\phi$ it is principally the amplitude of $Z_{\textrm{free}}$ which is important. The results of these tests can be summed up simply: while $M_b$ remains poorly constrained, $M_c$ can only rise above its zero eccentricity value in vary rare cases, and then not by much. The worst case we found was for $\phi=\pi/4$, where for $M_b=0$ the maximum mass for $M_c$ was $386M_\oplus$. The particular $\chi^2$ square surface varies for different input $\phi$ values. One example surface is shown in Figure \ref{figchisqphipiov4}, which also demonstrates an interesting degeneracy that arises between positive and negative $\triangle$ in that case.

Given the poorly mapped TTV curve, the sensitivity of our analysis to small corrections to the planetary periods, and the free eccentricity degeneracy it would be premature to make determinations of the planetary masses at this point. We can however limit both them and the corrections to the periods which would have to be made. As has been stated, $M_c < 350 M_\oplus$ in the zero free eccentricity case. At this point ($M_b=0$, $M_c=350M_\oplus$) we also obtain the maximum period correction to make to planet b. This is +0.029 days to $P_b$ in the zero free eccentricity case. As planet b's mass is less constrained placing a limit on the period correction for planet c is harder, but limiting $M_b$ to $600M_\oplus$ gives a maximum amplitude for the period correction to $P_c$ of -0.12 days. At the zero eccentricity case ($\phi=0$) we are already at the steepest gradient found on the TTV phase curve - as this sets the period correction, these amplitudes cannot go higher. They can however change sign (at $\phi=\pi$ for example), and so we constrain |Correction($P_b$)| < 0.029d and |Correction($P_c$)| < 0.12d immediately. The correction to $P_b$ can be further limited by noting that not all of the masses which provide a fit at $\phi=0$ do so at other values for $\phi$. In particular, when $\pi/2 < \phi < \pi$ (the case for a negative correction to $P_b$), the allowed range for $M_c$ is much smaller ($M_c\lesssim30M_\oplus$), which corresponds to a maximum negative correction of Correction($P_b$)>-0.002. As such, the final limits for $P_b$ are $-0.002 <$ Correction($P_b$) $< 0.029$, where Correction(P) is to be made to the periods found in Table \ref{tabparams}. This limits $-0.011 < \triangle < 0.013$ in the extreme case, confirming that the system remains very close to resonance.

At this stage it is worth stating the now better understood periods of these two objects. Using the period implied by our latest NITES transit measurement and the $T_0$ from \textit{K2}, we obtain $P_b = 7.921^{+0.028}_{-0.003}$ days , and $P_c = 11.91\pm 0.12$ days, where the errors are ranges rather than $1\sigma$ errors, and account for TTV related period corrections. \emph{When predicting transit times, these periods and the $T_0$ values of Table \ref{tabparams} should be used}. Note that there will also be possible TTVs of magnitude up to at least an hour.

\subsection{Hill stability}
\label{secthillstab}
Stable main sequence evolution of a two-planet system {\it may} be guaranteed by residing in a 3:2 MMR, although the Kirkwood gaps demonstrate that this MMR can instead harbour unstable orbits.  Also, as-yet-undetected planets perturbing the 3:2 MMR can cause complex dynamical structures \citep{fuse2002} and potentially instability.  The potential protection afforded by the 3:2 MMR becomes more important when the two planets are Hill unstable.  Two planets are Hill stable if their orbits are guaranteed to never cross; hence Neptune and Pluto are Hill unstable, but protected from each other by the 3:2 MMR.  Hill stability is a function of masses, semimajor axes, eccentricities and inclinations.  \cite{veretal2013} outline an algorithm for computing the Hill stability limit; no explicit formula exists for arbitrary eccentricities and inclinations.

In order for K2-19b \& c to be Hill unstable, their masses and/or eccentricities must be sufficiently large.  The mutual inclination between the planets of just about a degree negligibly affects the Hill stability limit \citep{verarm2004}\footnote{Just the existence of a nonzero mutual inclination between the planets indicates that the planets might instead reside in an inclination-based 6:4 MMR (see equation 1 of \citealt*{miletal1989}).}.  Constraining the eccentricities and masses based on orbital periods alone with Hill stability is a useful exercise but requires assumptions.   A commonly-made assumption for transiting planets is that those planets are on circular orbits; the closer the planet is to the star, the better that assumption, based on tidal circularisation arguments.  We need not make such assumptions here.

\begin{figure}
\resizebox{\hsize}{!}{\includegraphics{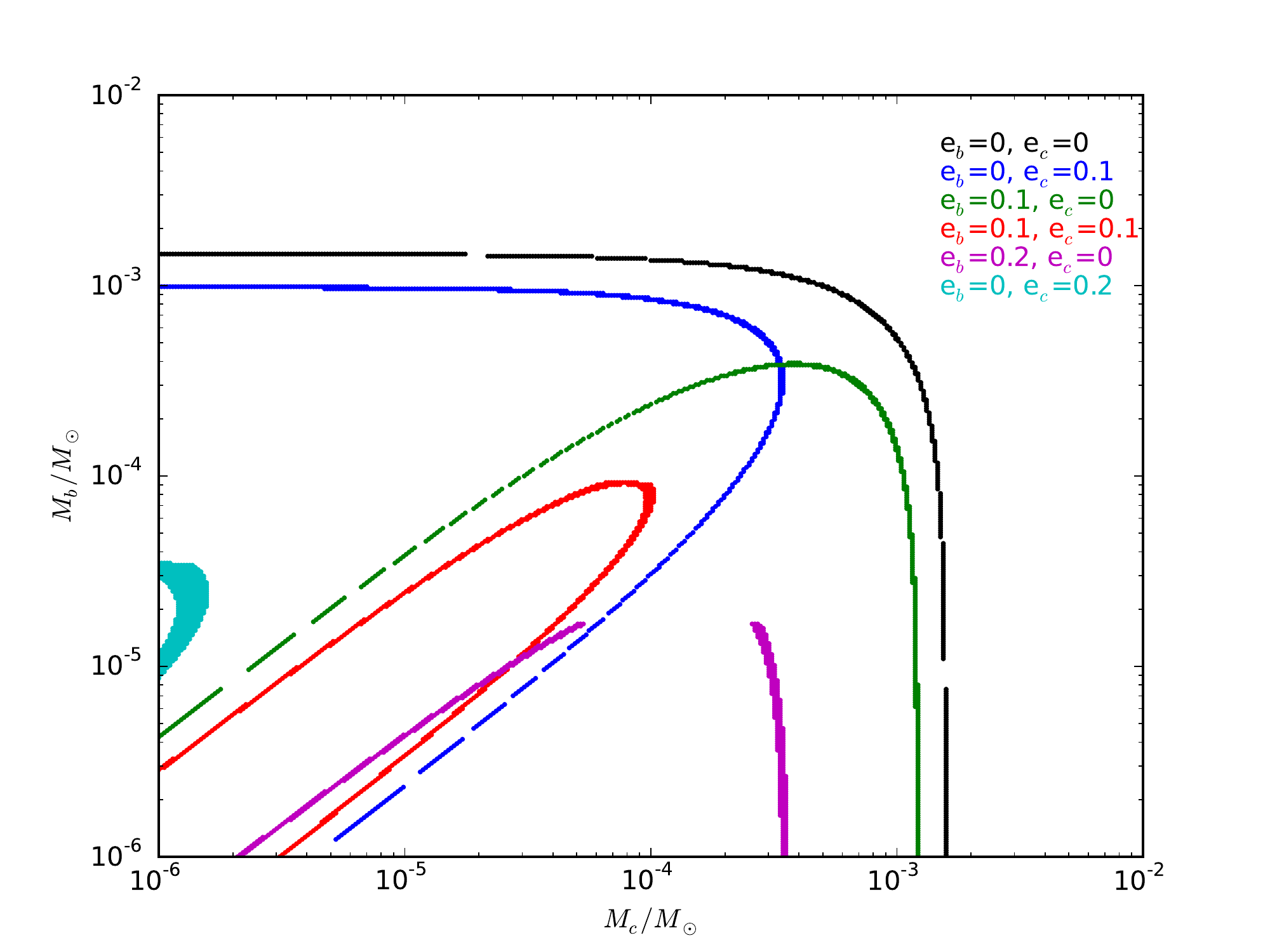}}
\caption{Hill stability limits for the K2-19 system.  We assume $M_{\star} = 0.9M_{\odot}$, the mutual inclination between the planets is 1 degree, and $a_{\rm b} / a_{\rm c}$ is within 0.1 per cent of $\left(3/2\right)^{2/3}$ based on the planetary orbital period ratios.  Eccentricites are measured in Jacobi coordinates.  If K2-19 is Hill unstable, then the 3:2 MMR might act as a crucial protection mechanism to ensure the system's long-term stability on the main sequence.}
\label{fighill}
\end{figure}

We can use the green curves in Fig. 1 of \cite{verfor2012} to roughly estimate Hill stability limits for K2-19.  Broadly, the plot shows that the system will be Hill stable if both (i) the sum of the eccentricities of both bodies (measured in Jacobi coordinates) does not exceed about 0.2, and (ii) that each planet is less massive than Jupiter.  These relations help motivate the setup for Figure \ref{fighill}.  The figure plots the pairs of planet masses for different eccentricities (measured in Jacobi coordinates) that would place the system on the edge of Hill stability, assuming a mutual inclination of 1 degree and a semimajor axis ratio which is just 0.1 per cent within $\left(3/2\right)^{2/3}$.  We generated each set of coloured points by sampling 600 different values of each of $M_{\rm b}$ and $M_{\rm c}$ uniformly in log space between $10^{-6} M_{\odot}$ and $10^{-2} M_{\odot}$, a range which covers both Earth masses and Jupiter masses.  The plot axes span the entire range of masses that we sampled.

Even if K2-19 is Hill stable, then planet c might eventually escape the system or planet b might crash into the star through Lagrange instability \citep{bargre2006,bargre2007,rayetal2009,kopbar2010,decetal2013,vermus2013}.  No analytical Lagrange unstable boundary is known to exist.  Regardless, the 3:2 MMR may then provide a protection mechanism not only for Hill unstable systems, but also for Lagrange unstable systems.  The upcoming space mission {\it PLATO} \citep{rauetal2014} will provide accurate enough stellar age constraints to potentially detect a decreasing trend in stable multi-planet systems with time due to Lagrange instability \citep{veretal2015}.

\section{Conclusion}
We have presented and validated K2-19b \& c, a system of two Neptune sized planets orbiting close to the 3:2 MMR, via transit observations with \textit{K2} and the NITES telescope and analysis using \texttt{PASTIS}. The inner, larger planet b shows high amplitude TTVs, and will likely show larger amplitudes when further transits are observed. The outer, smaller planet c can be expected to show even greater TTVs (scaled up by the mass ratio of the two planets), although these have not yet been observed. The precise ephemeris of these planets is still in doubt, see Sections \ref{sectlcfit} and \ref{sectdiscuss} for the fit values and limits to the possible corrections to them in the near but not in resonance case.

Future observations of K2-19 have the potential to lead to interesting discoveries. The system is bright enough to observe from the ground, leading to great potential for future work. The observation of more transits of either planet will lead to fully characterised TTV phase curves, as well as possibly being able to fully solve for the planetary masses via full dynamical analysis. Radial velocity observations have the potential to independently characterise the planetary masses, providing a window on the discrepancies that sometimes exist between radial velocity and TTV derived masses. We expect that this system is one of many more which will arise from the \textit{K2} mission.

\begin{acknowledgements}
We would like to thank the anonymous referee for their review of the manuscript. The data presented in this paper were obtained from the Mikulski Archive for Space Telescopes (MAST). STScI is operated by the Association of Universities for Research in Astronomy, Inc., under NASA contract NAS5-26555. Support for MAST for non-HST data is provided by the NASA Office of Space Science via grant NNX13AC07G and by other grants and contracts. A.S. is supported by the European Union under a Marie Curie Intra-European Fellowship for Career Development with reference FP7-PEOPLE-2013-IEF, number 627202. The authors would like to thank Thomas Marsh for use of his periodogram generating Python code. We acknowledge all the observers who attempt to detect the transits: Luc Arnold, Maurice Audejean, Mathieu Bachschmidt, Raoul Behrend, Laurent Bernasconi, the C2PU team, Jean-Christophe Dalouzy, Andr\'{e} Debackere, Serge Golovanow, Jo\~{a}o Gregorio, Patrick Martinez, Federico Manzini, Thierry Midavaine, Jacques Michelet, Romain Montaigut, G\'{e}rald Rousseau, Ren\'{e} Roy, Dominique Toublanc, Michael Vanhuysse, and Daniel Verilhac. SCCB thanks CNES for the grant 98761. O.D. thanks CNES for the grant 124378. DV benefited from support by the European Union through ERC Grant Number 320964.
\end{acknowledgements}





\bibliography{papers010315}

\begin{thebibliography}{40}
\expandafter\ifx\csname natexlab\endcsname\relax\def\natexlab#1{#1}\fi

\bibitem[{Agol {et~al.}(2005)Agol, Steffen, Sari, \& Clarkson}]{Agol:2005fp}
Agol, E., Steffen, J., Sari, R., \& Clarkson, W. 2005, Monthly Notices of the
  Royal Astronomical Society, 359, 567

\bibitem[Allard et al.(2012)]{2012IAUS..282..235A} Allard, F., Homeier, D., \& Freytag, B.\ 2012, IAU Symposium, 282, 235

\bibitem[Am{\^o}res \& L{\'e}pine(2005)]{2005AJ....130..659A} Am{\^o}res, E.~B., \& L{\'e}pine, J.~R.~D.\ 2005, \aj, 130, 659

\bibitem[{Armstrong {et~al.}(2015)Armstrong, Kirk, Lam, McCormac, Walker,
  Brown, Osborn, Pollacco, \& Spake}]{Armstrong:2015bn}
Armstrong, D.~J., Kirk, J., Lam, K. W.~F., {et~al.} 2015, Astronomy and
  Astrophysics, 579, A19

\bibitem[Barnes \& Greenberg(2006)]{bargre2006} Barnes, R., \& Greenberg, R.\ 2006, \apjl, 647, L163 

\bibitem[Barnes \& Greenberg(2007)]{bargre2007} Barnes, R., \& Greenberg, R.\ 2007, ApJL, 665, L67 

\bibitem[{{Baranne} {et~al.}(1996){Baranne}, {Queloz}, {Mayor}, {Adrianzyk},
  {Knispel}, {Kohler}, {Lacroix}, {Meunier}, {Rimbaud}, \&
  {Vin}}]{1996A&AS..119..373B}
{Baranne}, A., {Queloz}, D., {Mayor}, M., {et~al.} 1996, \aaps, 119, 373

\bibitem[{{Barros} {et~al.}(2013){Barros}, {Bou{\'e}}, {Gibson}, {Pollacco},
  {Santerne}, {Keenan}, {Skillen}, \& {Street}}]{2013MNRAS.430.3032B}
{Barros}, S.~C.~C., {Bou{\'e}}, G., {Gibson}, N.~P., {et~al.} 2013, \mnras,
  430, 3032

\bibitem[Batygin \& Morbidelli(2013)]{batmor2013} Batygin, K., \& Morbidelli, A.\ 2013, AJ, 145, 1 

\bibitem[{Beky {et~al.}(2014)Beky, Kipping, \& Holman}]{Beky:2014df}
Beky, B., Kipping, D.~M., \& Holman, M.~J. 2014, Monthly Notices of the Royal
  Astronomical Society, 442, 3686

\bibitem[{{Bouchy} {et~al.}(2009){Bouchy}, {H{\'e}brard}, {Udry}, {Delfosse},
  {Boisse}, {Desort}, {Bonfils}, {Eggenberger}, {Ehrenreich}, {Forveille},
  {Lagrange}, {Le Coroller}, {Lovis}, {Moutou}, {Pepe}, {Perrier}, {Pont},
  {Queloz}, {Santos}, {S{\'e}gransan}, \& {Vidal-Madjar}}]{2009A&A...505..853B}
{Bouchy}, F., {H{\'e}brard}, G., {Udry}, S., {et~al.} 2009, \aap, 505, 853

\bibitem[Callegari et al.(2006)]{caletal2006} Callegari, N., 
Ferraz-Mello, S., \& Michtchenko, T.~A.\ 2006, Celestial Mechanics and Dynamical Astronomy, 94, 381 

\bibitem[Chatterjee \& Ford(2014)]{chafor2014} Chatterjee, S., \& Ford, E.~B.\ 2015, ApJ, 803, 33 

\bibitem[Claret \& Bloemen(2011)]{2011A&A...529A..75C} Claret, A., \& Bloemen, S.\ 2011, \aap, 529, A75

\bibitem[{Crossfield {et~al.}(2015)Crossfield, Petigura, Schlieder, Howard,
  Fulton, Aller, Ciardi, Lepine, Barclay, de~Pater, de~Kleer, Quintana,
  Christiansen, Schlafly, Kaltenegger, Crepp, Henning, Obermeier, Deacon,
  Hansen, Liu, Greene, Howell, Barman, \& Mordasini}]{Crossfield:2015vm}
Crossfield, I. J.~M., Petigura, E., Schlieder, J., {et~al.}\ 2015, ApJ, 804, 10

\bibitem[{Deck \& Agol(2014)}]{Deck:2014wm}
Deck, K.~M. \& Agol, E. 2014, ApJ, 802, 116

\bibitem[Deck et al.(2013)]{decetal2013} Deck, K.~M., Payne, M., \& Holman, M.~J.\ 2013, ApJ, 774, 129 

\bibitem[Delisle et al.(2014)]{deletal2014} Delisle, J.-B., Laskar, J., \& Correia, A.~C.~M.\ 2014, A\&A, 566, AA137 

\bibitem[Delisle \& Laskar(2014)]{dellas2014} Delisle, J.-B., \& Laskar, J.\ 2014, A\&A, 570, LL7 

\bibitem[{{D{\'{\i}}az} {et~al.}(2014){D{\'{\i}}az}, {Almenara}, {Santerne},
  {Moutou}, {Lethuillier}, \& {Deleuil}}]{2014MNRAS.441..983D}
{D{\'{\i}}az}, R.~F., {Almenara}, J.~M., {Santerne}, A., {et~al.} 2014, \mnras,
  441, 983
  
\bibitem[Dotter et al.(2008)]{2008ApJS..178...89D} Dotter, A., Chaboyer, B., Jevremovi{\'c}, D., et al.\ 2008, \apjs, 178, 89

\bibitem[Emel'yanenko(2012)]{emel2012} Emel'yanenko, V.~V.\ 2012, Solar System Research, 46, 321 

\bibitem[Etzel(1981)]{1981psbs.conf..111E} Etzel, P.~B.\ 1981, Photometric and Spectroscopic Binary Systems, 111

\bibitem[{Everett {et~al.}(2015)Everett, Barclay, Ciardi, Horch, Howell, Crepp,
  \& Silva}]{Everett:2015fh}
Everett, M.~E., Barclay, T., Ciardi, D.~R., {et~al.} 2015, The Astronomical
  Journal, 149, 55

\bibitem[Fabrycky et al.(2014)]{fabetal2014} Fabrycky, D.~C., 
Lissauer, J.~J., Ragozzine, D., et al.\ 2014, ApJ, 790, 146 

\bibitem[{Foreman-Mackey {et~al.}(2015)Foreman-Mackey, Montet, Hogg, Morton,
  Wang, \& Sch{\"o}lkopf}]{ForemanMackey:2015vi}
Foreman-Mackey, D., Montet, B.~T., Hogg, D.~W., {et~al.} 2015, ApJ, 806, 215

\bibitem[Fuse(2002)]{fuse2002} Fuse, T.\ 2002, PASJ, 54, 493 

\bibitem[Go{\'z}dziewski et al.(2005)]{gozetal2005} 
Go{\'z}dziewski, K., Konacki, M., \& Wolszczan, A.\ 2005, ApJ, 619, 1084 

\bibitem[{Hadden \& Lithwick(2014)}]{Hadden:2014bf}
Hadden, S. \& Lithwick, Y. 2014, The Astrophysical Journal, 787, 80

\bibitem[Hadjidemetriou \& Voyatzis(2010)]{hadvoy2010} Hadjidemetriou, J.~D., \& Voyatzis, G.\ 2010, Celestial Mechanics and Dynamical Astronomy, 107, 3 

\bibitem[{Han {et~al.}(2014)Han, Wang, Wright, Feng, Zhao, Fakhouri, Brown, \&
  Hancock}]{Han:2014hn}
Han, E., Wang, S.~X., Wright, J.~T., {et~al.} 2014, Publications of the
  Astronomical Society of the Pacific, 126, 827

\bibitem[Henden et al.(2015)]{2015AAS...22533616H} Henden, A.~A., Levine, S., Terrell, D., \& Welch, D.~L.\ 2015, American Astronomical Society Meeting Abstracts, 225, \#336.16

\bibitem[{Holman(2005)}]{Holman:2005jf}
Holman, M.~J. 2005, Science, 307, 1288

\bibitem[Hormuth et al.(2008)]{2008JPhCS.131a2051H} Hormuth, F., Brandner, W., Hippler, S., \& Henning, T.\ 2008, Journal of Physics Conference Series, 131, 012051

\bibitem[{Howell {et~al.}(2014)Howell, Sobeck, Haas, Still, Barclay, Mullally,
  Troeltzsch, Aigrain, Bryson, Caldwell, Chaplin, Cochran, Huber, Marcy,
  Miglio, Najita, Smith, Twicken, \& Fortney}]{Howell:2014ju}
Howell, S.~B., Sobeck, C., Haas, M., {et~al.} 2014, Publications of the
  Astronomical Society of the Pacific, 126, 398

\bibitem[Kipping(2010)]{2010MNRAS.408.1758K} Kipping, D.~M.\ 2010, \mnras, 408, 1758

\bibitem[Kopparapu \& Barnes(2010)]{kopbar2010} Kopparapu, R.~K., \& Barnes, R.\ 2010, ApJ, 716, 1336 

\bibitem[Kordopatis et al.(2013)]{2013AJ....146..134K} Kordopatis, G., Gilmore, G., Steinmetz, M., et al.\ 2013, \aj, 146, 134

\bibitem[Kroupa(2001)]{2001MNRAS.322..231K} Kroupa, P.\ 2001, \mnras, 322, 231

\bibitem[{{Kurucz}(1993)}]{kurucz1993}
{Kurucz}, R. 1993, ATLAS9 Stellar Atmosphere Programs and 2 km/s grid.~Kurucz
  CD-ROM No.~13.~ Cambridge, Mass.: Smithsonian Astrophysical Observatory,
  1993., 13

\bibitem[{{Law} {et~al.}(2014){Law}, {Morton}, {Baranec}, {Riddle},
  {Ravichandran}, {Ziegler}, {Johnson}, {Tendulkar}, {Bui}, {Burse}, {Das},
  {Dekany}, {Kulkarni}, {Punnadi}, \& {Ramaprakash}}]{2014ApJ...791...35L}
{Law}, N.~M., {Morton}, T., {Baranec}, C., {et~al.} 2014, \apj, 791, 35


\bibitem[Lee et al.(2013)]{leeetal2013} Lee, M.~H., Fabrycky, D., 
\& Lin, D.~N.~C.\ 2013, ApJ, 774, 52


\bibitem[Libert \& Tsiganis(2011)]{libtsi2011} Libert, A.-S., \& Tsiganis, K.\ 2011, Celestial Mechanics and Dynamical Astronomy, 111, 201

\bibitem[{{Lillo-Box} {et~al.}(2012){Lillo-Box}, {Barrado}, \&
  {Bouy}}]{Lillo-boxMIT2012}
{Lillo-Box}, J., {Barrado}, D., \& {Bouy}, H. 2012, \aap, 546, A10

\bibitem[{{Lillo-Box} {et~al.}(2014){Lillo-Box}, {Barrado}, \&
  {Bouy}}]{Lillo-boxHRI2014}
{Lillo-Box}, J., {Barrado}, D., \& {Bouy}, H. 2014, \aap, 566, A103

\bibitem[Lissauer et al.(2011)]{lisetal2011} Lissauer, J.~J., 
Ragozzine, D., Fabrycky, D.~C., et al.\ 2011, ApJS, 197, 8 

\bibitem[{Lissauer {et~al.}(2012)Lissauer, Marcy, Rowe, Bryson, Adams,
  Buchhave, Ciardi, Cochran, Fabrycky, Ford, Fressin, Geary, Gilliland, Holman,
  Howell, Jenkins, Kinemuchi, Koch, Morehead, Ragozzine, Seader, Tanenbaum,
  Torres, \& Twicken}]{Lissauer:2012hq}
Lissauer, J.~J., Marcy, G.~W., Rowe, J.~F., {et~al.} 2012, The Astrophysical
  Journal, 750, 112

\bibitem[Lissauer et al.(2014)]{2014ApJ...784...44L} Lissauer, J.~J., Marcy, G.~W., Bryson, S.~T., et al.\ 2014, \apj, 784, 44

\bibitem[{Lithwick {et~al.}(2012)Lithwick, Xie, \& Wu}]{Lithwick:2012ud}
Lithwick, Y., Xie, J., \& Wu, Y. 2012, The Astrophysical Journal, 761, 122

\bibitem[{Lomb(1976)}]{Lomb:1976bo}
Lomb, N.~R. 1976, Astrophysics and Space Science, 39, 447

\bibitem[Malhotra et al.(1992)]{maletal1992} Malhotra, R., Black, D., Eck, A., \& Jackson, A.\ 1992, Nature, 356, 583 

\bibitem[Mayor et al.(2011)]{2011arXiv1109.2497M} Mayor, M., Marmier, M., Lovis, C., et al.\ 2011, arXiv:1109.2497

\bibitem[\protect\citeauthoryear{McCormac et al.}{2014}]{McCormac14} McCormac J., Skillen I., Pollacco D., et al. 2014, \mnras, 438, 3383

\bibitem[Milani et al.(1989)]{miletal1989} Milani, A., Nobili, A.~M., \& Carpino, M.\ 1989, Icarus, 82, 200

\bibitem[Morbidelli(2002)]{morby2002} Morbidelli, A.\ 2002, Modern celestial mechanics : aspects of solar system dynamics, by Alessandro Morbidelli.~London: Taylor Francis, 2002, ISBN 0415279399

\bibitem[Morton \& Johnson(2011)]{2011ApJ...738..170M} Morton, T.~D., \& Johnson, J.~A.\ 2011, \apj, 738, 170

\bibitem[Murray \& Dermott(1999)]{murder1999} Murray, C.~D., \& Dermott, S.~F.\ 1999, Solar system dynamics by Murray, C.~D., 1999, 

\bibitem[Nelson \& Davis(1972)]{1972ApJ...174..617N} Nelson, B., \& Davis, W.~D.\ 1972, \apj, 174, 617

\bibitem[{Nesvorn{\'y} {et~al.}(2014)Nesvorn{\'y}, Kipping, Terrell, \&
  Feroz}]{Nesvorny:2014ko}
Nesvorn{\'y}, D., Kipping, D., Terrell, D., \& Feroz, F. 2014, The
  Astrophysical Journal, 790, 31

\bibitem[{Nesvorn{\'y} \& Vokrouhlick{\'y}(2014)}]{Nesvorny:2014gm}
Nesvorn{\'y}, D. \& Vokrouhlick{\'y}, D. 2014, The Astrophysical Journal, 790,
  58

\bibitem[Ogihara \& Kobayashi(2013)]{ogikob2013} Ogihara, M., \& Kobayashi, H.\ 2013, ApJ, 775, 34 

\bibitem[Papaloizou \& Szuszkiewicz(2005)]{papszu2005} Papaloizou, J.~C.~B., \& Szuszkiewicz, E.\ 2005, MNRAS, 363, 153 

\bibitem[Peale \& Lee(2002)]{pealee2002} Peale, S.~J., \& Lee, M.~H.\ 2002, Science, 298, 593 

\bibitem[{{Pepe} {et~al.}(2002){Pepe}, {Mayor}, {Galland}, {Naef}, {Queloz},
  {Santos}, {Udry}, \& {Burnet}}]{2002A&A...388..632P}
{Pepe}, F., {Mayor}, M., {Galland}, F., {et~al.} 2002, \aap, 388, 632

\bibitem[{{Perruchot} {et~al.}(2008){Perruchot}, {Kohler}, {Bouchy}, {Richaud},
  {Richaud}, {Moreaux}, {Merzougui}, {Sottile}, {Hill}, {Knispel}, {Regal},
  {Meunier}, {Ilovaisky}, {Le Coroller}, {Gillet}, {Schmitt}, {Pepe}, {Fleury},
  {Sosnowska}, {Vors}, {M{\'e}gevand}, {Blanc}, {Carol}, {Point}, {Laloge}, \&
  {Brunel}}]{2008SPIE.7014E..0JP}
{Perruchot}, S., {Kohler}, D., {Bouchy}, F., {et~al.} 2008, in Society of
  Photo-Optical Instrumentation Engineers (SPIE) Conference Series, Vol. 7014,
  Society of Photo-Optical Instrumentation Engineers (SPIE) Conference Series,
  0

\bibitem[Petrovich et al.(2013)]{petetal2013} Petrovich, C., 
Malhotra, R., \& Tremaine, S.\ 2013, ApJ, 770, 24 

\bibitem[Pierens et al.(2014)]{pieetal2014} Pierens, A., Raymond, S.~N., Nesvorny, D., \& Morbidelli, A.\ 2014, ApJL, 795, LL11 

\bibitem[{Popper \& Etzel(1981)}]{Popper:1981de}
Popper, D.~M. \& Etzel, P.~B. 1981, The Astronomical Journal, 86, 102

\bibitem[{Press \& Rybicki(1989)}]{Press:1989hb}
Press, W.~H. \& Rybicki, G.~B. 1989, The Astrophysical Journal, 338, 277

\bibitem[Raghavan et al.(2010)]{2010ApJS..190....1R} Raghavan, D., McAlister, H.~A., Henry, T.~J., et al.\ 2010, \apjs, 190, 1

\bibitem[Rauer et al.(2014)]{rauetal2014} Rauer, H., Catala, C., Aerts, C., et al.\ 2014, Experimental Astronomy, 38, 249

\bibitem[Raymond et al.(2008)]{rayetal2008} Raymond, S.~N., Barnes, 
R., Armitage, P.~J., \& Gorelick, N.\ 2008, ApJL, 687, L107 

\bibitem[Raymond et al.(2009)]{rayetal2009} Raymond, S.~N., Barnes, R., Veras, D., et al.\ 2009, ApJL, 696, L98 

\bibitem[Rein et al.(2012)]{reietal2012} Rein, H., Payne, M.~J., 
Veras, D., \& Ford, E.~B.\ 2012, MNRAS, 426, 187 

\bibitem[Robin et al.(2003)]{2003A&A...409..523R} Robin, A.~C., Reyl{\'e}, C., Derri{\`e}re, S., \& Picaud, S.\ 2003, \aap, 409, 523

\bibitem[{Rowe {et~al.}(2014)Rowe, Bryson, Marcy, Lissauer, Jontof-Hutter,
  Mullally, Gilliland, Issacson, Ford, Howell, Borucki, Haas, Huber, Steffen,
  Thompson, Quintana, Barclay, Still, Fortney, Gautier, Hunter, Caldwell,
  Ciardi, DeVore, Cochran, Jenkins, Agol, Carter, \& Geary}]{Rowe:2014wz}
Rowe, J.~F., Bryson, S.~T., Marcy, G.~W., {et~al.} 2014, The Astrophysical
  Journal, 784, 45

\bibitem[{{Santerne} {et~al.}(2014){Santerne}, {H{\'e}brard}, {Deleuil},
  {Havel}, {Correia}, {Almenara}, {Alonso}, {Arnold}, {Barros}, {Behrend},
  {Bernasconi}, {Boisse}, {Bonomo}, {Bouchy}, {Bruno}, {Damiani},
  {D{\'{\i}}az}, {Gravallon}, {Guillot}, {Labrevoir}, {Montagnier}, {Moutou},
  {Rinner}, {Santos}, {Abe}, {Audejean}, {Bendjoya}, {Gillier}, {Gregorio},
  {Martinez}, {Michelet}, {Montaigut}, {Poncy}, {Rivet}, {Rousseau}, {Roy},
  {Suarez}, {Vanhuysse}, \& {Verilhac}}]{2014A&A...571A..37S}
{Santerne}, A., {H{\'e}brard}, G., {Deleuil}, M., {et~al.} 2014, \aap, 571, A37

\bibitem[Santerne et al.(2015)]{2015MNRAS.451.2337S} Santerne, A., D{\'{\i}}az, R.~F., Almenara, J.-M., et al.\ 2015, \mnras, 451, 2337

\bibitem[{Santos {et~al.}(2013)Santos, Sousa, Mortier, Neves, Adibekyan,
  Tsantaki, Delgado~Mena, Bonfils, Israelian, Mayor, \& Udry}]{Santos:2013dm}
Santos, N.~C., Sousa, S.~G., Mortier, A., {et~al.} 2013, Astronomy and
  Astrophysics, 556, A150

\bibitem[{Scargle(1982)}]{Scargle:1982eu}
Scargle, J.~D. 1982, The Astrophysical Journal, 263, 835

\bibitem[{{Sneden}(1973)}]{sneden1973}
{Sneden}, C.~A. 1973, PhD thesis, The University of Texas at Austin.

\bibitem[Snellgrove et al.(2001)]{sneetal2001} Snellgrove, M.~D., Papaloizou, J.~C.~B., \& Nelson, R.~P.\ 2001, A\&A, 374, 1092 

\bibitem[Southworth(2008)]{2008MNRAS.386.1644S} Southworth, J.\ 2008, \mnras, 386, 1644

\bibitem[{Southworth(2013)}]{Southworth:2013gc}
Southworth, J. 2013, Astronomy and Astrophysics, 557, A119

\bibitem[{Steffen {et~al.}(2012)Steffen, Fabrycky, Ford, Carter, Desert,
  Fressin, Holman, Lissauer, Moorhead, Rowe, Ragozzine, Welsh, Batalha,
  Borucki, Buchhave, Bryson, Caldwell, Charbonneau, Ciardi, Cochran, Endl,
  Everett, Gautier, Gilliland, Girouard, Jenkins, Horch, Howell, Isaacson,
  Klaus, Koch, Latham, Li, Lucas, MacQueen, Marcy, McCauliff, Middour, Morris,
  Mullally, Quinn, Quintana, Shporer, Still, Tenenbaum, Thompson, Twicken, \&
  Van~Cleve}]{Steffen:2012kx}
Steffen, J.~H., Fabrycky, D.~C., Ford, E.~B., {et~al.} 2012, Monthly Notices of
  the Royal Astronomical Society, 421, 2342

\bibitem[\protect\citeauthoryear{Stetson}{1987}]{1987PASP...99..191S} Stetson P.~B., 1987, PASP, 99, 191

\bibitem[Strehl(1902)]{1902AN....158...89S} Strehl, K.\ 1902, Astronomische Nachrichten, 158, 89

\bibitem[Szab{\'o} et al.(2011)]{2011ApJ...736L...4S} Szab{\'o}, G.~M., Szab{\'o}, R., Benk{\H o}, J.~M., et al.\ 2011, \apjl, 736, LL4

\bibitem[Tadeu dos Santos et al.(2015)]{tadetal2015} Tadeu dos Santos, M., Correa-Otto, J.~A., Michtchenko, T.~A., \& Ferraz-Mello, S.\ 2015, A\&A, 573, AA94

\bibitem[Torres et al.(2005)]{2005ApJ...619..558T} Torres, G., Konacki, M., Sasselov, D.~D., \& Jha, S.\ 2005, \apj, 619, 558

\bibitem[{Torres {et~al.}(2010{\natexlab{a}})Torres, Andersen, \&
  Gim{\'e}nez}]{Torres:2010eoa}
Torres, G., Andersen, J., \& Gim{\'e}nez, A. 2010{\natexlab{a}}, The Astronomy
  and Astrophysics Review, 18, 67

\bibitem[{Torres {et~al.}(2010{\natexlab{b}})Torres, Fressin, Batalha, Borucki,
  Brown, Bryson, Buchhave, Charbonneau, Ciardi, Dunham, Fabrycky, Ford,
  Gautier, Gilliland, Holman, Howell, Isaacson, Jenkins, Koch, Latham,
  Lissauer, Marcy, Monet, Prsa, Quinn, Ragozzine, Rowe, Sasselov, Steffen, \&
  Welsh}]{Torres:2010eob}
Torres, G., Fressin, F., Batalha, N.~M., {et~al.} 2010{\natexlab{b}}, The
  Astrophysical Journal, 727, 24

\bibitem[{{Tsantaki} {et~al.}(2013){Tsantaki}, {Sousa}, {Adibekyan}, {Santos},
  {Mortier}, \& {Israelian}}]{tsantaki2013}
{Tsantaki}, M., {Sousa}, S.~G., {Adibekyan}, V.~Z., {et~al.} 2013, \aap, 555,
  A150

\bibitem[Tuomi \& Jones(2012)]{2012A&A...544A.116T} Tuomi, M., \& Jones, H.~R.~A.\ 2012, \aap, 544, A116

\bibitem[{Vanderburg \& Johnson(2014)}]{Vanderburg:2014bi}
Vanderburg, A. \& Johnson, J.~A. 2014, Publications of the Astronomical Society
  of the Pacific, 126, 948

\bibitem[{Vanderburg {et~al.}(2014)Vanderburg, Montet, Johnson, Buchhave, Zeng,
  Pepe, Collier~Cameron, Latham, Molinari, Udry, Lovis, Matthews, Cameron, Law,
  Bowler, Angus, Baranec, Bieryla, Boschin, Charbonneau, Cosentino, Dumusque,
  Figueira, Guenther, Harutyunyan, Hellier, Kuschnig, Lopez-Morales, Mayor,
  Micela, Moffat, Pedani, Phillips, Piotto, Pollacco, Queloz, Rice, Riddle,
  Rowe, Rucinski, Sasselov, Segransan, Sozzetti, Szentgyorgyi, Watson, \&
  Weiss}]{Vanderburg:2014wi}
Vanderburg, A., Montet, B.~T., Johnson, J.~A., {et~al.} 2014, ApJ, 800, 59

\bibitem[Veras \& Armitage(2004)]{verarm2004} Veras, D., \& Armitage, P.~J.\ 2004, Icarus, 172, 349 

\bibitem[Veras \& Ford(2012)]{verfor2012} Veras, D., \& Ford, E.~B.\ 2012, MNRAS, 420, L23 

\bibitem[Veras et al.(2013)]{veretal2013} Veras, D., Mustill, A.~J., Bonsor, A., \& Wyatt, M.~C.\ 2013, MNRAS, 431, 1686 

\bibitem[Veras \& Mustill(2013)]{vermus2013} Veras, D., \& Mustill, A.~J.\ 2013, MNRAS, 434, L11 

\bibitem[Veras et al.(2015)]{veretal2015} Veras, D., Brown, D.~J.~A., Mustill, A.~J., Pollacco, D.\ 2015, accepted to MNRAS

\bibitem[Wang et al.(2014)]{2014ApJ...791..111W} Wang, J., Fischer, D.~A., Xie, J.-W., \& Ciardi, D.~R.\ 2014, \apj, 791, 111

\bibitem[Walsh et al.(2011)]{waletal2011} Walsh, K.~J., Morbidelli, 
A., Raymond, S.~N., O'Brien, D.~P., \& Mandell, A.~M.\ 2011, Nature, 475, 206 

\bibitem[Wolszczan \& Frail(1992)]{wolfra1992} Wolszczan, A., \& Frail, D.~A.\ 1992, Nature, 355, 145 

\bibitem[Wolszczan(1994)]{wolszczan1994} Wolszczan, A.\ 1994, 
Science, 264, 538 

\bibitem[Wang \& Ji(2014)]{wanji2014} Wang, S., \& Ji, J.\ 2014, ApJ, 795, 85

\bibitem[Wright et al.(2010)]{2010AJ....140.1868W} Wright, E.~L., Eisenhardt, P.~R.~M., Mainzer, A.~K., et al.\ 2010, \aj, 140, 1868

\bibitem[{Xie(2014)}]{Xie:2014jk}
Xie, J.-W. 2014, The Astrophysical Journal Supplement Series, 210, 25

\bibitem[Zhang et al.(2014)]{zhaetal2014} Zhang, X., Li, H., Li, 
S., \& Lin, D.~N.~C.\ 2014, ApJL, 789, LL23 

\end{thebibliography}
\bibliographystyle{aa}

\appendix

\onecolumn
\section{\texttt{PASTIS} supplements}

\begin{figure*}[h]
\begin{center}
\resizebox{\hsize}{!}{\includegraphics{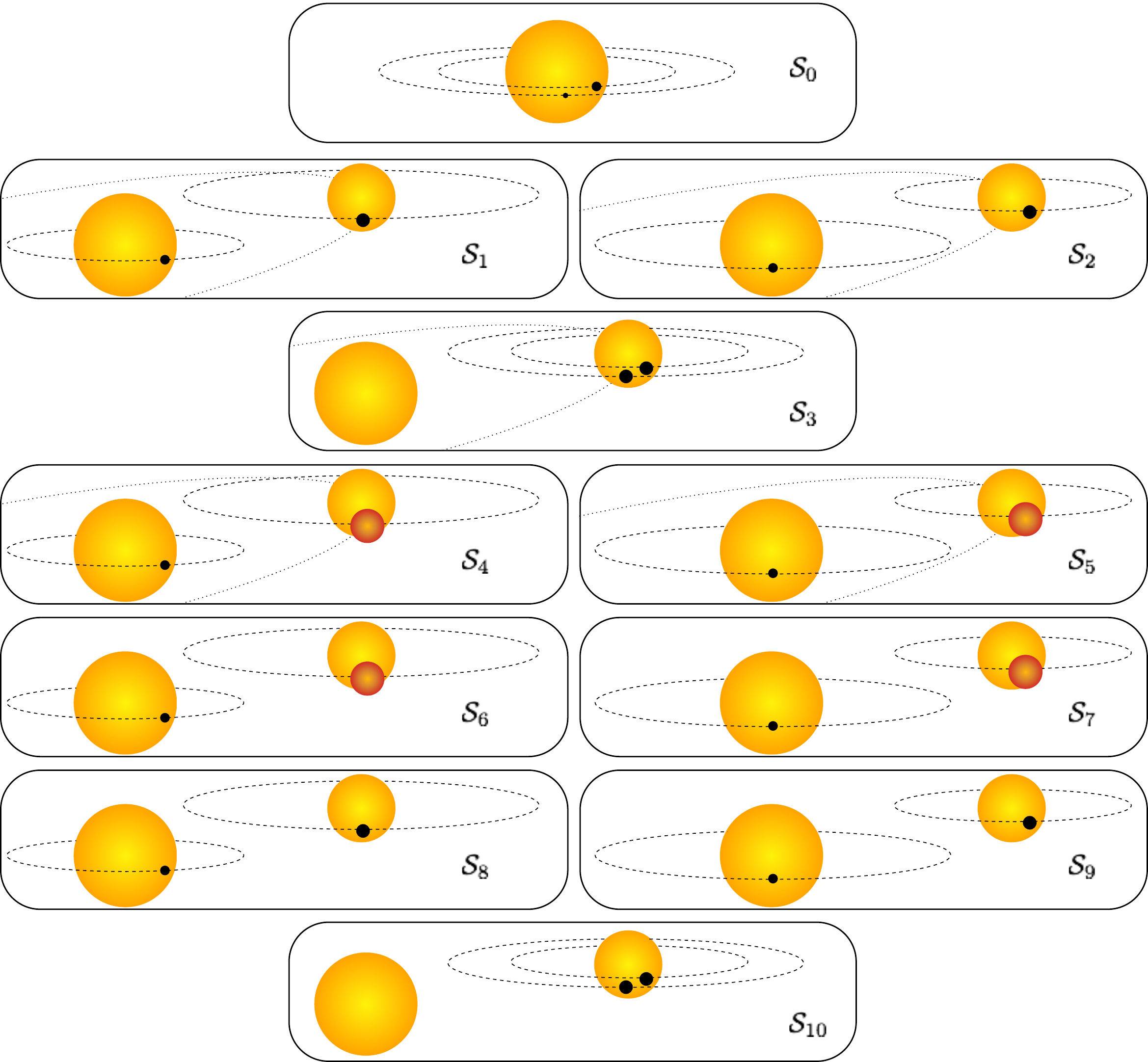}}
\caption{Sketchs of the different scenarios considered for the \texttt{PASTIS} validation of the two planets in the K2-19 system. The sizes are not proportional to their real values but note the relative size of the orbits between the 8-d and 12-d signals. Note also the orbit of the contaminant system in scenarios $\mathcal{S}_1$ to $\mathcal{S}_5$, which marks the difference between these and the unbound scenarios $\mathcal{S}_6$ to $\mathcal{S}_{10}$.}
\label{Sketchs}
\end{center}
\end{figure*}

\begin{landscape}
\begin{longtable}{l*{11}{>{\centering\arraybackslash}p{1.4cm}}}
\caption{Priors used in the \texttt{PASTIS} analyses: $\mathcal{U}(a,b)$ represents a Uniform prior between $a$ and $b$; $\mathcal{N}(\mu,\sigma^{2})$ represents a Normal distribution with a mean of $\mu$ and a width of $\sigma^{2}$; $\mathcal{P}(\alpha; x_{min}; x_{max})$ represents a Power Law distribution with an exponent $\alpha$ computed between $x_{min}$ and $x_{max}$ ; $\mathcal{P}_{2}(\alpha_{1}; \alpha_{2}; x_{0}; x_{min}; x_{max})$ represents a double Power Law distribution with an exponent $\alpha_{1}$ computed between $x_{min}$ and $x_{0}$ and an exponent $\alpha_{2}$ computed between $x_{0}$ and $x_{max}$; and finally $\mathcal{S}(a,b)$ represents a Sine distribution between $a$ and $b$.}\\
\hline
\hline
Parameter & $\mathcal{S}_{0}$ & $\mathcal{S}_{1}$ & $\mathcal{S}_{2}$ & $\mathcal{S}_{3}$ & $\mathcal{S}_{4}$ & $\mathcal{S}_{5}$ & $\mathcal{S}_{6}$ & $\mathcal{S}_{7}$ & $\mathcal{S}_{8}$ & $\mathcal{S}_{9}$ & $\mathcal{S}_{10}$ \\
\hline
\endfirsthead
\caption{Continued.} \\
\hline
Parameter & $\mathcal{S}_{0}$ & $\mathcal{S}_{1}$ & $\mathcal{S}_{2}$ & $\mathcal{S}_{3}$ & $\mathcal{S}_{4}$ & $\mathcal{S}_{5}$ & $\mathcal{S}_{6}$ & $\mathcal{S}_{7}$ & $\mathcal{S}_{8}$ & $\mathcal{S}_{9}$ & $\mathcal{S}_{10}$ \\
\hline
\endhead
\hline
\endfoot
\hline
\hline
\endlastfoot
\multicolumn{12}{l}{\textit{Target star parameters}}\\
& & & \\
Effective temperature T$_{\rm eff}$ [K] & \multicolumn{11}{c}{$\dotfill~\mathcal{N}(5230; 417)~\dotfill$} \\
Surface gravity $\log g$ [cm.s$^{-2}$] & \multicolumn{11}{c}{$\dotfill~\mathcal{N}(4.47; 0.83)~\dotfill$} \\
Iron abondance [Fe/H] [dex] & \multicolumn{11}{c}{$\dotfill~\mathcal{N}(0.38; 0.23)~\dotfill$} \\
Distance $d_{t}$ [pc] & \multicolumn{11}{c}{$\dotfill~\mathcal{P}(2.0; 10; 10000)~\dotfill$}\\
& & & \\
\hline
\multicolumn{12}{l}{\textit{Primary blend star parameters}}\\
& & & \\
Stellar mass M$_{\star_{1}}$ [$\mathrm{M}_{\astrosun}$] & n/a & \multicolumn{10}{c}{$\dotfill~\mathcal{P}_{2}(-1.3; -2.3; 0.5; 0.1; 2)~\dotfill$} \\
Iron abondance [Fe/H]$_{\star_{1}}$ [dex] & n/a & n/a & n/a & n/a & n/a & n/a & \multicolumn{5}{c}{$\dotfill~\mathcal{U}(-2.5; 0.5)~\dotfill$} \\
Age $\tau_{\star_{1}}$ [Gyr] & n/a & n/a & n/a & n/a & n/a & n/a & \multicolumn{5}{c}{$\dotfill~\mathcal{U}(0; 13.8)~\dotfill$} \\
Distance $d_{\star_{1}}$ [pc] & n/a & n/a & n/a & n/a & n/a & n/a & \multicolumn{5}{c}{$\dotfill~\mathcal{P}(2.0; 10; 10000)~\dotfill$}\\
& & & \\
\hline
\multicolumn{12}{l}{\textit{Secondary blend star parameters}}\\
& & & \\
Stellar mass M$_{\star_{2}}$ [$\mathrm{M}_{\astrosun}$] & n/a & n/a & n/a & n/a & \multicolumn{4}{c}{$\dotfill~\mathcal{P}_{2}(-1.3; -2.3; 0.5; 0.1; 2)~\dotfill$} & n/a & n/a & n/a\\
& & & \\
\hline
\multicolumn{12}{l}{\textit{8-d planet parameters}}\\
& & & \\
Planetary radius R$_{p}^{b}$ [R$_{jup}$] & \multicolumn{5}{c}{$\dotfill~\mathcal{U}(0; 2.2)~\dotfill$} & n/a &  $\mathcal{U}(0; 2.2)$ & n/a & \multicolumn{3}{c}{$\dotfill~\mathcal{U}(0; 2.2)~\dotfill$}\\
& & & \\
\hline
\multicolumn{12}{l}{\textit{12-d planet parameters}}\\
& & & \\
Planetary radius R$_{p}^{c}$ [R$_{jup}$] & \multicolumn{4}{c}{$\dotfill~\mathcal{U}(0; 2.2)~\dotfill$} & n/a & $\mathcal{U}(0; 2.2)$ & n/a &  \multicolumn{4}{c}{$\dotfill~\mathcal{U}(0; 2.2)~\dotfill$}\\
& & & \\
\hline
\multicolumn{12}{l}{\textit{8-d orbit parameters}}\\
& & & \\
Orbital period $P^{b}$ [d] & \multicolumn{11}{c}{$\dotfill~\mathcal{N}(7.919454; 8\times10^{-3})~\dotfill$}\\
Epoch $T_{0}^{b}$ [BJD - 2450000] & \multicolumn{11}{c}{$\dotfill~\mathcal{N}(6813.38345; 0.03)~\dotfill$}\\
Orbital inclination $i^{b}$ [\degr] & \multicolumn{11}{c}{$\dotfill~\mathcal{S}(70; 90)~\dotfill$}\\
& & & \\
\hline
\multicolumn{12}{l}{\textit{12-d orbit parameters}}\\
& & & \\
Orbital period $P^{c}$ [d] & \multicolumn{11}{c}{$\dotfill~\mathcal{N}(11.90701; 4\times10^{-2})~\dotfill$}\\
Epoch $T_{0}^{c}$ [BJD - 2450000] & \multicolumn{11}{c}{$\dotfill~\mathcal{N}(6817.2759; 0.1)~\dotfill$}\\
Orbital inclination $i^{c}$ [\degr] & \multicolumn{11}{c}{$\dotfill~\mathcal{S}(70; 90)~\dotfill$}\\
& & & \\
\newpage
\multicolumn{12}{l}{\textit{\textit{Kepler} data parameters}}\\
& & & \\
Contamination & \multicolumn{11}{c}{$\dotfill~\mathcal{N}(0; 0.02)~\dotfill$} \\
Out-of-transit flux & \multicolumn{11}{c}{$\dotfill~\mathcal{U}(0.9; 1.1)~\dotfill$} \\
Jitter & \multicolumn{11}{c}{$\dotfill~\mathcal{U}(0; 0.1)~\dotfill$}\\
& & & \\
\hline
\multicolumn{12}{l}{\textit{Spectral Energy Distribution parameter}}\\
& & & \\
Jitter [mag] & \multicolumn{11}{c}{$\dotfill~\mathcal{U}(0; 1)~\dotfill$}\\
& & & \\
\end{longtable}%
\label{PASTISpriors}
\end{landscape}

\begin{landscape}
\begin{longtable}{l*{11}{>{\centering\arraybackslash}p{1.4cm}}}
\caption{Posterior distributions of the \texttt{PASTIS} analyses.}\\
\hline
\hline
Parameter & $\mathcal{S}_{0}$ & $\mathcal{S}_{1}$ & $\mathcal{S}_{2}$ & $\mathcal{S}_{3}$ & $\mathcal{S}_{4}$ & $\mathcal{S}_{5}$ & $\mathcal{S}_{6}$ & $\mathcal{S}_{7}$ & $\mathcal{S}_{8}$ & $\mathcal{S}_{9}$ & $\mathcal{S}_{10}$ \\
\hline
\endfirsthead
\caption{Continued.} \\
\hline
Parameter & $\mathcal{S}_{0}$ & $\mathcal{S}_{1}$ & $\mathcal{S}_{2}$ & $\mathcal{S}_{3}$ & $\mathcal{S}_{4}$ & $\mathcal{S}_{5}$ & $\mathcal{S}_{6}$ & $\mathcal{S}_{7}$ & $\mathcal{S}_{8}$ & $\mathcal{S}_{9}$ & $\mathcal{S}_{10}$ \\
\hline
\endhead
\hline
\endfoot
\hline
\hline
\endlastfoot
\multicolumn{12}{l}{\textit{Target star parameters}}\\
& & & \\
Effective temperature T$_{\rm eff}$ [K] & 5562$\pm$41 & 5774$\pm$51 & 5756$\pm$53 & 5732$\pm$58 & 5550$\pm$130 & 5752$\pm$55 & 5487$\pm$61 & 5308$\pm$98 & 5490$\pm$66 & 5333$\pm$210 & 5368$\pm$84\\
Surface gravity $\log g$ [cm.s$^{-2}$] & 4.509 $^{_{+0.026}}_{^{-0.051}}$ & 4.397 $\pm$0.036 & 4.426 $^{_{+0.033}}_{^{-0.052}}$ & 4.358 $\pm$0.057 & 4.527 $^{_{+0.037}}_{^{-0.061}}$ & 4.332 $^{_{+0.033}}_{^{-0.050}}$ & 4.508 $\pm$0.059 & 4.27 $\pm$0.20 & 4.508 $\pm$0.058 & 4.441 $^{_{+0.072}}_{^{-0.190}}$ & 2.88 $\pm$0.35 \\
Iron abondance [Fe/H] [dex] & 0.07$\pm$0.16 & 0.24$\pm$0.17 & 0.03$\pm$0.18 & 0.03$\pm$0.20 & $-0.08\pm$0.15 & 0.14$\pm$0.17 & 0.10$\pm$0.19 & 0.33$^{_{+0.11}}_{^{-0.18}}$ & 0.08$\pm$0.19 & 0.29$\pm$0.18 & 0.38$^{_{+0.09}}_{^{-0.16}}$ \\
Distance $d_{t}$ [pc] & 329$^{_{+24}}_{^{-15}}$ & 501$\pm$18 & 470$\pm$19 & 499$\pm$21 & 452$\pm$23 & 521$\pm$24 & 329$\pm$25 & 443 $\pm$ 96 & 329$\pm$27 & 730$^{_{+480}}_{^{-250}}$ & 3800$\pm$1900 \\
& & & \\
\hline
\multicolumn{12}{l}{\textit{Primary blend star parameters}}\\
& & & \\
Stellar mass M$_{\star_{1}}$ [$\mathrm{M}_{\astrosun}$] & n/a & 0.914 $\pm$0.048 & 0.862 $\pm$0.059 & 0.858 $\pm$0.058 & 0.265 $\pm$0.026 & 0.883 $\pm$0.048 & 1.16 $\pm$0.18 & 0.933 $^{_{+0.110}}_{^{-0.068}}$ & 1.00\ \ \ \ \ \ $^{_{+0.26}}_{^{-0.15}}$ & 0.824 $\pm$0.079 & 0.90 $\pm$0.11\\
Iron abondance [Fe/H]$_{\star_{1}}$ [dex] & n/a & n/a & n/a & n/a & n/a & n/a & -1.8$\pm$0.5 & -1.5$\pm$0.7 & -1.8$^{_{+0.8}}_{^{-0.5}}$ & -0.6$^{_{+0.3}}_{^{-0.5}}$ & -1.8$\pm$0.6 \\
Age $\tau_{\star_{1}}$ [Gyr] & n/a & n/a & n/a & n/a & n/a & n/a & 1.19$^{_{+1.6}}_{^{-0.76}}$ & 1.3$^{_{+2.4}}_{^{-1.0}}$ & 1.5$^{_{+3.8}}_{^{-1.1}}$ & 5.6$\pm$4.4 & 2.1$^{_{+4.4}}_{^{-1.6}}$ \\
Distance $d_{\star_{1}}$ [pc] & n/a & n/a & n/a & n/a & n/a & n/a & 3200 $\pm$1300 & 820\ \ \ \ \ \ $^{_{+320}}_{^{-120}}$ & 2260\ \ \ \ \ \  $^{_{+1300}}_{^{-720}}$ & 336\ \ \ \ \ \  $^{_{+190}}_{^{-39}}$ & 880\ \ \ \ \ \  $^{_{+240}}_{^{-150}}$\\
& & & \\
\hline
\multicolumn{12}{l}{\textit{Secondary blend star parameters}}\\
& & & \\
Stellar mass M$_{\star_{2}}$ [$\mathrm{M}_{\astrosun}$] & n/a & n/a & n/a & n/a & 0.928 $\pm$0.056 & 0.103 $^{_{+0.005}}_{^{-0.002}}$ & 0.186 $\pm$0.064 & 0.112 $^{_{+0.016}}_{^{-0.009}}$ & n/a & n/a & n/a\\
& & & \\
\hline
\multicolumn{12}{l}{\textit{8-d planet parameters}}\\
& & & \\
Planetary radius R$_{p}^{b}$ [R$_\textrm{Jup}$] & 0.661 $^{_{+0.053}}_{^{-0.033}}$ & 0.960 $\pm$0.040 & 1.087 $\pm$0.053 & 1.141 $\pm$0.050 & 0.906 $\pm$0.037 & n/a &  0.677 $\pm$0.060 & n/a & 0.678 $\pm$0.063 & 0.661 $^{_{+0.26}}_{^{-0.069}}$ & 1.268 $\pm$0.12\\
& & & \\
\hline
\multicolumn{12}{l}{\textit{12-d planet parameters}}\\
& & & \\
Planetary radius R$_{p}^{c}$ [R$_\textrm{Jup}$] & 0.385 $^{_{+0.031}}_{^{-0.018}}$ & 0.679 $\pm$0.029 & 0.521 $^{_{+0.025}}_{^{-0.016}}$ & 0.669 $\pm$0.028 & n/a & 0.580 $^{_{+0.037}}_{^{-0.021}}$ & n/a & 0.60 $\pm$0.17 & 1.65 $\pm$0.32 & 0.99\ \ \ \ \ \ \  $^{_{+0.55}}_{^{-0.33}}$ & 0.743 $\pm$0.070\\
& & & \\
\hline
\multicolumn{12}{l}{\textit{8-d orbit parameters}}\\
& & & \\
Orbital period $P^{b}$ [d] & 7.919491 $\pm$8.3$\times10^{-5}$ & 7.919510 $\pm$9.0$\times10^{-5}$ & 7.919527 $\pm$9.0$\times10^{-5}$ & 7.919525 $\pm$9.1$\times10^{-5}$ & 7.919502 $\pm$8.6$\times10^{-5}$ & 7.919520 $\pm$9.0$\times10^{-5}$ & 7.919494 $\pm$8.3$\times10^{-5}$ & 7.919486 $\pm$8.6$\times10^{-5}$ & 7.919486 $\pm$8.7$\times10^{-5}$ & 7.919477 $\pm$8.7$\times10^{-5}$ & 7.919501 $\pm$9.0$\times10^{-5}$\\
Epoch $T_{0}^{b}$ [BJD - 2456000] & 813.38351 $\pm$3.8$\times10^{-4}$ & 813.38347 $\pm$4.1$\times10^{-4}$ & 813.38340 $\pm$3.9$\times10^{-4}$ & 813.38341 $\pm$4.0$\times10^{-4}$ & 813.38348 $\pm$4.1$\times10^{-4}$ & 813.38345 $\pm$4.2$\times10^{-4}$ & 813.38350 $\pm$3.8$\times10^{-4}$ & 813.38358 $\pm$4.0$\times10^{-4}$ & 813.38351 $\pm$3.8$\times10^{-4}$ & 813.38355 $\pm$3.9$\times10^{-4}$ & 813.38351 $\pm$4.1$\times10^{-4}$\\
Orbital inclination $i^{b}$ [\degr] & 88.7$^{_{+0.3}}_{^{-0.4}}$ & 87.8$\pm$0.2 & 89.7$\pm$0.3 & 89.1$\pm$0.3 & 88.9$\pm$0.6 & 89.5$\pm$0.4 & 88.7$\pm$0.5 & 89.4$\pm$0.5 & 88.7$\pm$0.5 & 89.4$\pm$0.5 & 89.1$\pm$0.2\\
& & & \\
\newpage
\multicolumn{12}{l}{\textit{12-d orbit parameters}}\\
& & & \\
Orbital period $P^{c}$ [d] & 11.90664 $\pm$4.9$\times10^{-4}$ & 11.90634 $\pm$4.9$\times10^{-4}$ & 11.90661 $\pm$4.7$\times10^{-4}$ & 11.90642 $\pm$5.0$\times10^{-4}$ & 11.90636 $\pm$5.1$\times10^{-4}$ & 11.90730 $\pm$5.0$\times10^{-4}$ & 11.90650 $\pm$5.2$\times10^{-4}$ & 11.90721 $\pm$4.9$\times10^{-4}$ & 11.90650 $\pm$4.9$\times10^{-4}$ & 11.90659 $\pm$5.1$\times10^{-4}$ & 11.90637 $\pm$5.0$\times10^{-4}$ \\
Epoch $T_{0}^{c}$ [BJD - 2456000] & 817.2757 $\pm$1.4$\times10^{-3}$ & 817.2758 $\pm$1.4$\times10^{-3}$ & 817.2758 $\pm$1.4$\times10^{-3}$ & 817.2758 $\pm$1.4$\times10^{-3}$ & 817.2760 $\pm$1.5$\times10^{-3}$ & 817.2735 $\pm$1.5$\times10^{-3}$ & 817.2760 $\pm$1.5$\times10^{-3}$ & 817.2739 $\pm$1.5$\times10^{-3}$ & 817.2760 $\pm$1.5$\times10^{-3}$ & 817.2760 $\pm$1.4$\times10^{-3}$ & 817.2758 $\pm$1.5$\times10^{-3}$\\
Orbital inclination $i^{c}$ [\degr] & 89.4$\pm$0.5 & 89.8$^{_{+0.1}}_{^{-0.2}}$ & 88.7$^{_{+0.2}}_{^{-0.3}}$ & 89.8$\pm$0.2 & 89.8$\pm$0.2 & 88.1$\pm$0.2 & 89.4$\pm$0.6 & 87.9$\pm$1.2 & 89.4$\pm$0.6 & 88.9$^{_{+0.8}}_{^{-0.1}}$ & 89.7$\pm$0.3\\
& & & \\
\hline
\multicolumn{12}{l}{\textit{\textit{Kepler} data parameters}}\\
& & & \\
Contamination & -0.001 $\pm$0.021 & -0.003 $\pm$0.020 & -0.003 $\pm$0.020 & -0.004 $\pm$0.021 & -0.002 $\pm$0.020 & 0.001 $\pm$0.021 & -0.002 $\pm$0.021 & -0.001 $\pm$0.021 & -0.001 $\pm$0.020 & -0.001 $\pm$0.020 & -0.002 $\pm$0.020\\
Out-of-transit flux & 1.0000223 $\pm$9.0$\times10^{-6}$ & 1.0000228 $\pm$9.2$\times10^{-6}$ & 1.0000236 $\pm$9.6$\times10^{-6}$ & 1.0000238 $\pm$9.4$\times10^{-6}$ & 1.0000101 $\pm$9.2$\times10^{-6}$ & 1.0000320 $\pm$9.3$\times10^{-6}$ & 1.0000228 $\pm$9.3$\times10^{-6}$ & 1.0000315 $\pm$9.9$\times10^{-6}$ & 1.0000234 $\pm$9.4$\times10^{-6}$ & 1.0000221 $\pm$8.9$\times10^{-6}$ & 1.0000230 $\pm$9.5$\times10^{-6}$\\
Jitter [ppm] & 59$\pm$23 & 63$\pm$21 & 67$\pm$20 & 68$\pm$19 & 68$\pm$20 & 68$\pm$20 & 60$\pm$23 & 60$\pm$22 & 60$\pm$23 & 58$\pm$24 & 65$\pm$22 \\
& & & \\
\hline
\multicolumn{12}{l}{\textit{Spectral Energy Distribution parameter}}\\
& & & \\
Jitter [mag] & 0.04$^{_{+0.02}}_{^{-0.1}}$ & 0.04$^{_{+0.02}}_{^{-0.1}}$ & 0.04$^{_{+0.02}}_{^{-0.01}}$ & 0.04$^{_{+0.02}}_{^{-0.01}}$ & 0.04$^{_{+0.02}}_{^{-0.01}}$ & 0.04$^{_{+0.02}}_{^{-0.01}}$ & 0.04$^{_{+0.02}}_{^{-0.01}}$ & 0.04$^{_{+0.02}}_{^{-0.01}}$ & 0.04$^{_{+0.02}}_{^{-0.01}}$ & 0.04$^{_{+0.02}}_{^{-0.01}}$ & 0.04$^{_{+0.02}}_{^{-0.01}}$ \\
& & & \\
\end{longtable}%
\label{PASTISresults}
\end{landscape}

\end{document}